\documentclass[preprint,12pt,authoryear]{elsarticle}
\usepackage[a4paper, total={6.5in, 9.5in}]{geometry}
\usepackage{amssymb}
\usepackage{amsmath}
\usepackage{wrapfig}
\usepackage[version=3]{mhchem}
\usepackage{tabularx,booktabs}
\usepackage{float}
\usepackage[colorlinks=true,citecolor=cyan,linkcolor=green]{hyperref}
\usepackage[capitalize,noabbrev]{cleveref}
\usepackage{booktabs}
\usepackage{longtable}
\usepackage[dvipsnames]{xcolor ,colortbl}
\usepackage{lineno}
\usepackage{multirow}
\usepackage{setspace}
\graphicspath{{figures/}}

\usepackage[title]{appendix}
\DeclareUnicodeCharacter{2212}{-}

\usepackage[export]{adjustbox}

\usepackage{pdflscape}
\bibpunct[]{(}{)}{;}{a}{,}{,}

\journal{Nuclear Physics B}
\begin{document}

\begin{frontmatter}

\title{Earth's composition: origin, evolution and energy budget}

\author[a,b]{William F. McDonough}

\affiliation[a]{Advanced Institute for Marine Ecosystem Change, Department of Earth Sciences and Research Center for Neutrino Science, Tohoku University, Sendai, Miyagi 980-8578, Japan}

\affiliation[b]{Department of Geology, University of Maryland, College Park, MD 20742, USA }

\begin{abstract}
One in every two atoms in the Earth, Mars, and the Moon is oxygen; it is the third most abundant element in the solar system. The oxygen isotopic compositions of the terrestrial planets are different from those of the Sun and demonstrate that these planets are not direct compositional analogs of the solar photosphere. Likewise, the Sun's O/Fe, Fe/Mg and Mg/Si values are distinct from those of inner solar system chondrites and terrestrial planets. These four elements (O, Fe, Mg, Si) make up 90\% to 94\% by mass (and atomic \%) of the rocky planets and their abundances are determined uniquely using geophysical, geochemical and cosmochemical constraints. 

The rocky planets grew rapidly ($<$10 million years) from large populations of planetesimals, most of which were differentiated, having a core and a mantle, before being accreted. Planetary growth in the early stages of protoplanetary disk evolution was rapid and was only partially recorded by the meteoritic record. The noncarbonaceous meteorites (NC) provide insights into the early history of the inner solar system and are used to construct a framework for how the rocky planets were assembled. NC chondrites have chondrule ages that are two to three million years younger than $t_{zero}$ (the age of calcium-aluminum inclusions, CAI), documenting that chondrites are middle- to late-stage products of solar system evolution. 

The composition of the Earth, its current form of mantle convection, and the amount of radiogenic power that drives its engine remain controversial topics. Earth’s dynamics are driven by primordial and radiogenic heat sources. Measurement of the Earth's geoneutrino flux defines its radiogenic power and restricts its bulk composition. Using the latest data from the KamLAND and Borexino geoneutrino experiments affirms that the Earth has $\leq$20 TW of radiogenic power and sets the proportions of refractory lithophile elements in the bulk silicate Earth at 2.5 to 2.7 times that in CI chondrites. The bulk Earth and the bulk Mars are enriched in refractory elements about 1.9 times that of the CI chondrites. Earth is more volatile-depleted and less oxidized than Mars. 
\end{abstract}
 
\begin{keyword}
BSE (bulk silicate Earth) \sep  pyrolite \sep  heat production \sep geoneutrino
\footnote{wfm@tohoku.ac.jp, mcdonoug@umd.edu;  ORCID number: 0000-0001-9154-3673\\ This paper is dedicated to my mentor, colleague, and good friend, Shen-su Sun}
\end{keyword}

\end{frontmatter}

\section{Introduction}

Earth and other rocky planets have undergone a unique physical, chemical, and isotopic evolution compared to the gas giants that are sourced in the outer solar system. It is not clear whether our solar system history is typical and therefore directly applicable to understanding the origin, growth, and evolution of exoplanetary systems. However, it serves to characterize the major evolutionary stages of our inner rocky planets, particularly as we begin to understand the causes and consequences of the contrasting growth environment of the inner versus outer solar system, specifically, the rocky planets versus the gas giants. 

Meteorites include stony meteorites (chondrites and achondrites), stony irons, and irons. Chondrites are undifferentiated mixtures of silicates (chondrules, matrix, and calcium-aluminum inclusions) and metals, and so are the most primitive. They are the closest analogs to rocky planets, as they contain both metal and silicate components. These meteorites provide the bases for establishing the age, origin, and processes involved in planet and solar system construction. Some also contain abundant organic compounds, which can provide clues to the origin of life \citep{pizzarello2006chemistry}. Achondrites represent a wide range of meteorites derived from differentiated planets and planetesimals and are mainly made up of silicates. Iron meteorites are iron-nickel alloys mostly from the cores of planetesimals.

Orbiting the Sun and located between Mars and Jupiter, the asteroid belt represents a notional boundary between the inner and outer solar system. The outer portions of the asteroid belt are dominated by C-type asteroids that are rich in organic compounds and likely chemically and isotopically similar to carbonaceous chondrite meteorites (CC), presumed samples from the outer solar system \citep{kurokawa2022distant}. In contrast, the inner asteroid belt has more S-type asteroids that are associated with noncarbonaceous chondrite meteorites (NC), mostly ordinary chondrites (OC) from the inner solar system \citep{de2022composition,burkhardt2021terrestrial,kleine2020non}. [See Section \ref{bias} for more details.]

However, no chondrite has a composition equal to that of the Earth or other terrestrial planet, and none can be simply identified as approaching the composition of the Earth without ad hoc modifications of its composition to make the two bodies match \citep{javoy2010chemical,dauphas2017isotopic}. The meteoritic record contains incomplete information on their parent bodies and is mostly restricted to samples delivered to Earth in the last 10$^6$ years. Moreover, the presence of Kirkwood gaps in the asteroid belt, the lack of achondrites that can be genetically linked with iron meteorite parent bodies, and the large sample population for ordinary chondrites (15 to 30 times greater than carbonaceous and enstatite chondrites) bias our perspective.

Despite the caveats mentioned above, chondrites provide a record of chemical and petrological processes that occurred in the protoplanetary disk, in parent bodies, and in chondrule-forming and storage environments. Likewise, nonchondritic meteorites provide insight into parent body processes, and together all samples reveal a rich temporal and spatial record of the origins, accretion, and evolution of the rocky planets. To model the origins of the rocky planets, I use the half-mass condensation temperature scale (T$_{50}$ in K at 10 Pa of H and C/O $\sim$0.5) of \cite{lodders2003solar} and the behavior of elements in terms of lithophile, siderophile and / or chalcophile as defined by \cite{goldschmidt1937principles} and summarized in \cite{fischer2025earth}. The behavior of refractory elements (T$_{50} >$ 1350 K; e.g., Ca, Al, Mo, W), main elements (1290 K $> T_{50} <$ 1350 K; Mg, Si, Fe, Ni), and moderately volatile elements (600 K $> T_{50} <$ 1290 K; e.g., alkali metals, S, and Pb) are increasingly being understood from this meteoritic record. However, the behavior of volatile elements ($T_{50} <$ 600 K; e.g., the ices and noble gases) is less well established.

Understanding the origin and growth of the rocky planets continues to be a focus of discussion.  Here I attempt to explain the origin and early evolution of the Earth, and by analogy all of the rocky planets and samples from the asteroid belt. Meteorites provide insights into the early history of the inner solar system and are used to construct a framework for how the rocky planets were assembled. Given the 30 years since an earlier version of this paper was published \citep{mcdonough1995}, many new insights have been recognized, vast amounts of chemical and isotopic data have been published, and geoneutrino studies have begun, all of which have guided our re-interpretations of Earth's story. 

\section{Biases in the meteoritic and cometary record} \label{bias}

There is a compositional dichotomy between meteorite groups: non-carbonaceous (NC) and carbonaceous chondrite (CC) (both groups include chondrites and non-chondritic meteorites) \citep{trinquier2007widespread,trinquier2009origin,warren2011stable,kruijer2017age}. NC chondrites are generally more reduced than CC chondrites, with the exception that the R group of chondrites is oxidized in the NC group and CK being a reduced member of the CC group. Ordinary, enstatite and R chondrites are NC types and are interpreted to originate in the inner solar system, while CC chondrites are interpreted to originate from the outer solar system \citep{warren2011stable,kruijer2017age}. Earth and NC chondrites have compositional links \citep{javoy2010chemical,dauphas2017isotopic}.

To properly interpret the meteoritic and cometary record, one needs to understand the nature of its intrinsic biases. Importantly, there are materials that are not in our meteorite collections. Meteorites are derived from a range of localities, including the Apollo objects (i.e., Near-Earth asteroids), the main asteroid belt, and the Kuiper belt. There are many biases that influence our meteoritic record (see \underline{The Meteoritical Bulletin} \url{https://www.lpi.usra.edu/meteor/}):
\begin{itemize}
    \item Kirkwood gaps in the main asteroid belt document radial zones of selective mass depletions; different radial zones in the asteroid belt typically have a predominance of certain asteroid types.
    \item The meteoritic record has been provided for the last 10$^6$ years, except for rare examples from ancient sedimentary rocks \citep{kyte1998meteorite}, however, these are rare and are often concluded to be comparable in composition to meteorites in our modern inventory.
    \item Of the 77,500 meteorites identified, only $<$ 1,400 are identified and recovered as Falls (meteorites observed to fall to Earth, limiting exposure and risk of terrestrial contamination), with most falling in the last 1100 years. However, all meteorites are altered during their flight through the atmosphere and even during their brief residence time on the Earth’s surface before recovery.

    \item Statistics for the $\sim$2\% of Falls are: 80\% ordinary chondrites, $<$5\% carbonaceous chondrites, and $<$2\% enstatite chondrites (EC). The statistics for $\sim$98\% of Finds are much more biased towards achondrites and iron meteorites.
    \item Most Falls appear to have come from only a few significant asteroidal belt collisions. Two major impacts some $\sim$40 and 466 million years ago may be responsible for a significant bias in ordinary chondrites in our collection \citep{brovz2024young,marsset2024massalia}.
    \item Based on Falls, there are $\sim$1000 ordinary chondrites and $<$100 carbonaceous and  enstatite chondrites combined; therefore, our statistical evaluations of chondrites are limited.
    \item Calcium-Aluminum Rich Inclusions (CAIs) are high-temperature condensates of a gas of solar composition and are refractory in nature. CAIs are the first solids formed in the solar system that are preserved and sampled. Their age \citep{amelin2002lead,bouvier2010age,connelly2012absolute}, 4567.3 $\pm$ 0.2 Ma (with a systematic uncertainty of about $\pm$ 1 Ma), and brief formation duration, a few 10$^4$ to a few 10$^5$ a \citep{kawasaki2020variations}, defines the age of the solar system. These refractory inclusions record the highest temperatures of condensates and thus are probably mainly formed in the innermost portion of the solar nebula near the Sun. However, it is unusual that they are a common phase in carbonaceous chondrites and rare in NC chondrites \citep{dunham2023calcium}. This preservation bias might reflect the incorporation of inner solar system CAIs into the rocky planets during their formation, leaving only leftovers for the NC chondrites to inherit. 
    \item We do not have clear examples of the silicate mantles associated with iron meteorite parent bodies. Primitive achondrites, winonaites, lodranites, acapulcoites, ureilites, and brachinites, along with pallasites, are all relatively rare, mostly formed during the first few million years post-CAI formation. These materials offer limited insight into the silicate shells surrounding the NC metallic cores.  
    \item Gas-solid condensation, density sorting, and magnetic sorting are unidirectional, nonisochemical processes that produce compositional differences between the planets and the protoplanetary disk.
\end{itemize}

The meteoritic record comes mostly from the main asteroid belt, which in contrast to the planets, is populated with early solar system debris having eccentric and inclined orbits that promote collisions and result in disruption rather than accretion. In general, the inner part of the belt is dominated by S-type asteroids (that is, NC-like) and the outer part by C-type (carbonaceous; CC-like) asteroids \citep{demeo2014solar,demeo2015compositional}, which is consistent with their regional origins.

The first and most obvious bias in our meteorite record is illustrated by the existence of Kirkwood gaps in the main asteroid belt; these are zones in the asteroid belt depleted in asteroids. The positions of Kirkwood gaps in the asteroid belt are matched by mean motion resonances (MMRs) and two significant secular resonances (SRs). These perturbations are due to the orbital period of an asteroid being an integer multiple of Jupiter’s orbital period (MMR) or when an asteroid and Saturn precession frequencies are proportional. This gravitational garden process of the asteroid belt has been ongoing for $>$4.5 billion years.  Prominent Kirkwood gaps are found at the mean motion resonances of 4:1, 3:1, 5:2 and 2:1. The current total mass of the asteroid belt is about 40,000 times less than Earth's mass (Table \ref{table:T1}). The initial mass of the asteroid belt has been widely debated, with estimates ranging from a factor of two of its current size to several earth masses (see the review in \cite{clement2024formation}).

\begin{table} 
\centering
\caption{Physical properties of solid bodies in the inner solar system and the Sun }
\begin{tabular}{r|ccccc}
\\[-2ex]
\hline
 & diameter   & mass       & Earth mass     & uncompressed    & distance from \\
bodies     &(km)   & (10$^{24}$ kg) & fraction & density (kg/m$^3$) & Sun (AU)        \\ 
\hline & \\[-2ex]
Sun        & 1,391,400  & 1.989 $\times 10^6$   & 330,000$\times$  & 1410   & 0       \\
Mercury    & 4879  & 0.3301       & 5.5\%    & 5300  & 0.39          \\
Venus      & 12,104  & 4.867       & 82\%     & 4400  & 0.72          \\
Earth      & 12,756  & 5.972       & 100\%    & 4400  & 1.0 \\
Moon       & 3475  & 0.07346      & 1.2\%    & 3300  & 1.0 \\
Mars       & 6792  & 0.6417      & 11\%     & 3800  & 1.52          \\
asteroid belt$^*$ &  m to 1000 km        & 0.0024     & 0.04\%   & 1800-3500       & 1.5-5         \\
big asteroids$^\dagger$ & 12\% to 27\% lunar$^\ddagger$   & 0.00149   & 0.025\%   & 2100-3500       & 2.4-3.1       \\
\hline

\end{tabular}

\par\medskip \raggedright
$^*$ There are between 1 and 2 million asteroids $>$1 km in diameter in the main asteroid belt.

$^\dagger$ mass sum of 4 biggest asteroids (1 Ceres, 4 Vesta, 2 Pallas, and  10 Hygiea). 

$^\ddagger$diameter expressed in relative lunar diameter.
\label{table:T1}
\end{table}

Many nucleosynthetic isotopic systems are used to distinguish NC from CC meteorites \citep{warren2011stable} and document the isotopic dichotomy \citep{bermingham2020nc,kruijer2020great}. Importantly, these genetic isotopic distinctions are useful tracers of mixing components within and between the NC and CC reservoirs. Increasingly, they show, for example, that the building blocks of the rocky planets come predominantly from the NC, inner solar system sources \citep{dauphas2017isotopic}, with the final 10 to 20\% by weight of Earth's accretion, including impact material from the moon-forming event(s), also coming predominantly from NC materials \citep{bermingham2024non}, although some studies have argued for comparatively higher percentages of CC materials during the final major stage of accretion \citep{budde2019molybdenum}.

These known biases present us with challenges in reconstructing the early history of the solar system and assessing the representativeness of meteorites as the building blocks of the planets. However, many technological advances have been made bringing new insights into the nature and origin of chondrites and nonchondrites. In addition, new samples (e.g., Ryugu, Bennu, Erg Chech 002 (EC002)) have brought new insights. For example, EC002 is an andesite, it is the oldest (4.565 Ma) known silica rich crustal rock and it was derived from a differentiated protoplanet that was accreted some 4,566 Myrs ago \citep{barrat20214}, 2.3 Myrs after CAI formation.

\section{The protoplanetary disk (PPD): lifetime and evolution }

Known as the birthplaces of planets, protoplanetary disks (PPD) have a lifetime ranging from 1 to 30 million years, with typical lifetimes of 2-3 million years depending on the size of the star \citep{li2016lifetimes}. A PPD is a rotating gas and dust cloud with a composition comparable to that of the its star and formed via the gravitational collapse of a molecular cloud. Their evolution is regulated by a balance of gravitational, magnetic, gas pressure, and rotational forces. The termination of a PPD is defined as when it loses its gas and dust. In some cases, disk lifetimes can last up to 30 million years \citep{long2025first}. Overall, the evolution of a PPD and its lifetime depend on the mass of the system star, as well as the dispersal rate of the disk during its evolution through accretion, magnetohydrodynamic winds, and photoevaporation \citep{komaki2023simulations}. Our protoplanetary disk probably existed for some 3 to 5 million years, more or less \citep{li2016lifetimes,borlina2022lifetime}. Unfortunately, from astronomical insights, we cannot constrain the PPD lifetime for our solar system to be better than a factor of two, although an apparent inverse relationship exists between stellar mass PPD lifetimes (e.g. 2-3 Myr) \citep{komabayashi2014thermodynamics}. 

Chondrules and Ryugu particles are being used as recorders of the early solar system
magnetic field as a function of the heliocentric position \citep{weiss2021history,maurel20244,mansbach2024evidence,borlina2022lifetime}. These data provide potential insights into an evolving PPD and inform our models of its evolution through the magnetohydrodynamic wind stage. Completing the process and slowing planetary growth is the final stage of PPD evolution, photoevaporation, where high-energy photon radiation (UV and X-rays) ionizes gases, causing them to move outward in a heliocentric dispersal. The amount of mass that the Earth grew during and after the loss of the disk is unknown. Progressively, one expects continued accretion of chondritic-like materials, but given the lower temperatures and loss of the enveloping hydrogen-rich protoplanetary disk, later planetary growth was likely slower, cooler, and more oxidized.

Chondrites, particularly NC types, are later-stage residual products of an evolving PPD. Although chondrule formation might have occurred from the time of CAI formation to the end of the existence of PPD, it seems that most NC chondrules formed some two to three million years after CAI formation \citep{pape2019time,connolly2016chondrules,siron2022high}.

The collapsing and rotating molecular cloud was accompanied by rapid inflow of material into the star and outflow of material to conserve angular momentum \citep{williams2011protoplanetary}. Frictional heating of the disk, particularly on the high density midplane during these processes, produced a heliocentric temperature gradient. Here, I focus on the compositional attributes of the condensing phases.

\begin{wrapfigure}{R}{0.5\textwidth}
\centering
\includegraphics[width=.90\linewidth]{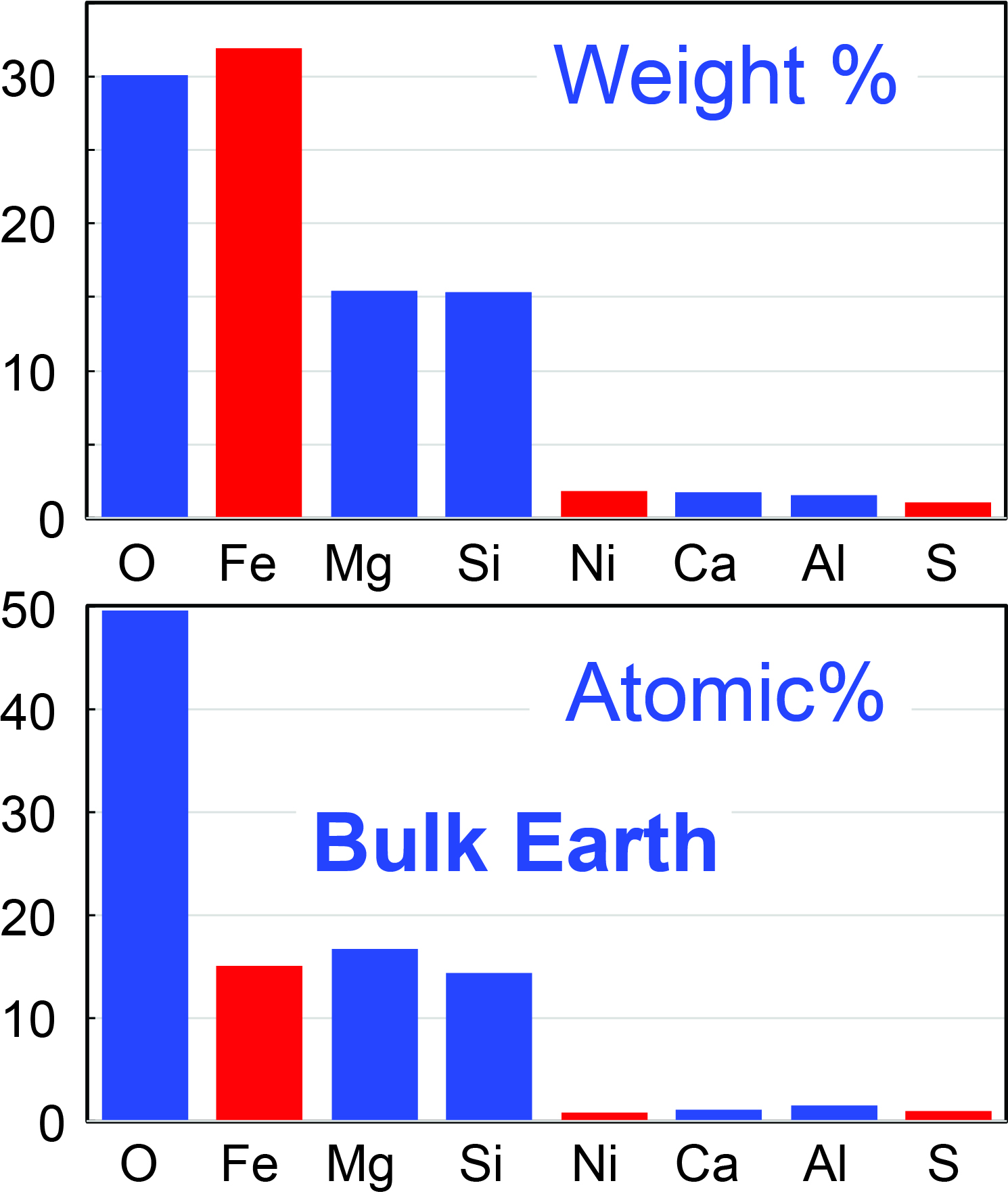}
\caption{The weight and atomic proportions of the 8 most abundant elements in the Earth and that probably in the other rocky planets and the Moon \citep{mcdonough2021tp}. These eight elements typically total $\geq$99 wt\% of a planet's mass. Red elements are concentrated in the core.}
\label{fig:8elem}
\end{wrapfigure}

The protoplanetary disk evolves temporally and spatially by experiencing heating and then cooling, with the latter stage accompanied by the condensation of minerals in a temperature-dependent sequence, as recorded in chondritic minerals \citep{grossman1972condensation}. [Note that in the early days of the solar system, time and space dimensions are poorly resolved.] Mineral condensation and incorporation into growing planetesimals results in non-isochemical evolution of planets relative to the PPD, which explains the difference in the oxygen isotopic compositions of the planets and meteorites and of the Sun \citep{mckeegan2011oxygen}. In terms of the four main elements that compose the terrestrial planets (O, Mg, Si, Fe), oxygen is the most abundant, making up one out of every two atoms in these planets (Figure \ref{fig:8elem}). Importantly, none of these condensation and accretion processes affects the overall composition of PPD as their removal from the disk represents an insignificant loss of mass (see Table \ref{table:T1}).

The early evolution of the PPD starts from a hot inner solar system (i.e. source of NC materials) dominated by condensates of olivine and Fe-Ni alloys that give rise to planetesimals and planets having Mg-rich silicates surrounding metallic cores. During accretion of differentiated planetesimals and planets, differences in the physical properties of metals versus silicates (i.e., mechanical strength, magnetism, and thermal conductivity) probably control some of the final compositions of the planets (e.g., O/Fe) \citep{mcdonough2021tp}.

The condensation of the four main elements (O, Mg, Si and Fe), plus Al, Ca, and Ni, which comprise some $\sim$98 to 99 wt\% (and atomic\%) of the rocky planets (Figure \ref{fig:8elem}), does not require the incorporation of any moderately volatile elements. The addition of these latter three elements assumes that Ca/Al and Fe/Ni ratios are fixed (1.07 $\pm$ 0.04 and 17.4 $\pm$ 0.5, respectively) using constant chondritic ratios. Incorporation of elements with lower $T_{50}$ condensation temperatures depends on the evolving heliocentric temperature gradient of the PPD.  In this regard, the contrast between the CC and NC chondrites is significant.  CC chondrites show a compositional dependency, in which those with more matrix and fewer chondrules have less depletion in moderately volatile elements, both lithophile and siderophile \citep{braukmuller2019earth,braukmuller2018chemical,braukmuller2025moderately}. By contrast, NC chondrites have a much higher proportion of chondrules and limited amounts of matrix. These chondrites are depleted in moderately volatile siderophile elements compared to lithophilic elements.  This uniquely inner solar system signature is likely due to condensation of those more evolved, lower-temperature silicates effectively incorporating more of the moderately volatile lithophile elements.

\begin{figure*}[h]
\centering
\includegraphics[width=0.95\linewidth]{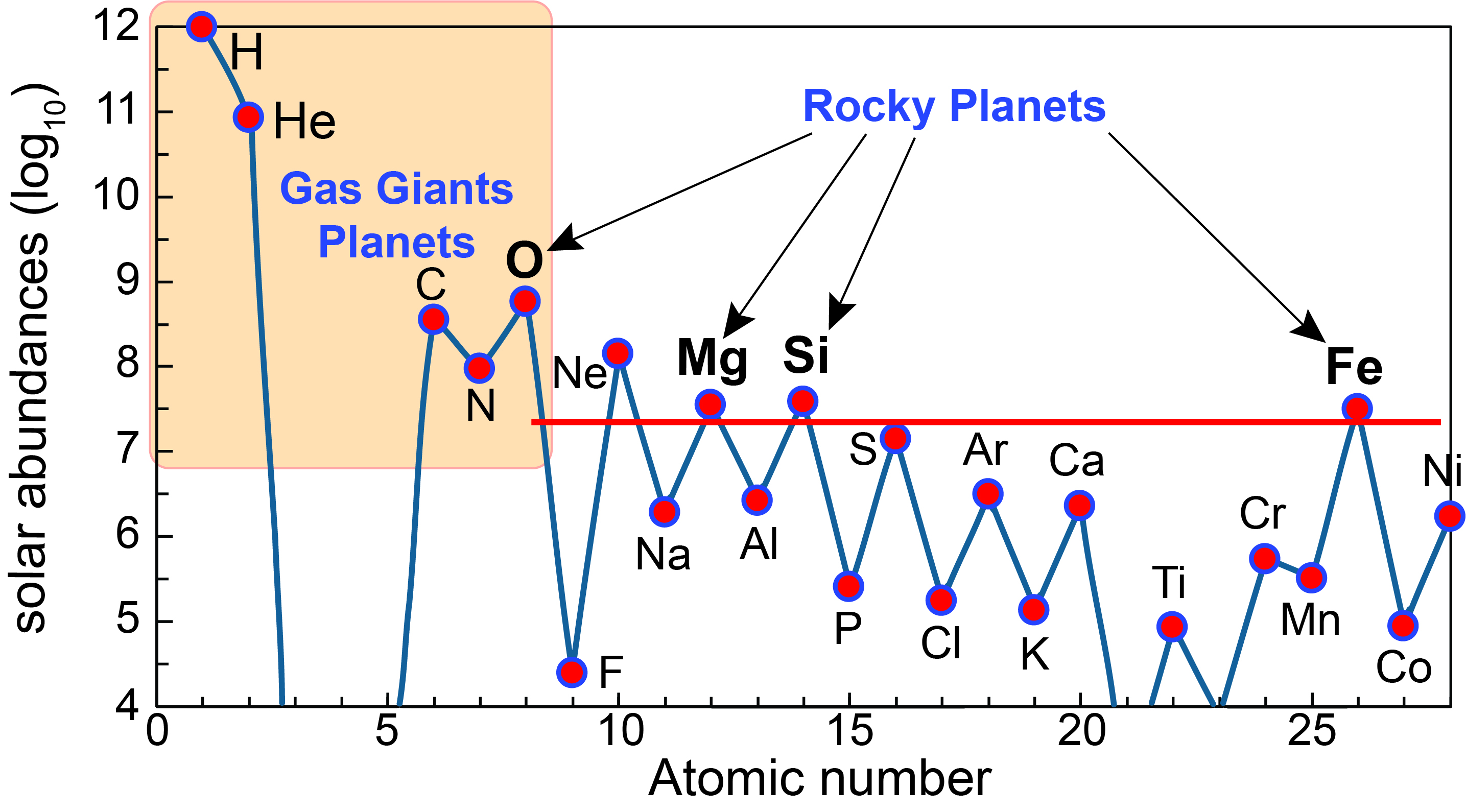}
\caption{The solar chemical abundance curve, with element abundance expressed as A$_i$ = log$_{10}$(N$_i$/N$_H$) + 12, where the number of atoms of N$_i$ is plotted according to its atomic number. Data are from \cite{asplund2009chemical,magg2022observational}. The most abundant elements in the gas and ice giants (H, He, C, N, and O) differ from those in the rocky planets (O, Fe, Mg, and Si).}
\label{fig:elem-abund}
\end{figure*}

\section{Inner solar system growth and evolution of the rocky planets} 

With planet construction starting in the early fiery days of protoplanetary disk evolution, Earth and other rocky planets were made from building blocks that had slightly different 'DNA' than those available to gas giants, as evidenced by the isotopic distinctions of NC versus CC meteorites \citep{warren2011stable,kruijer2017age}. The parent bodies of the NC and CC iron meteorites were built and differentiated rapidly, albeit somewhat slower for the CC bodies \citep{hilton2022chemical}; abundant small-body accretion later gave way to the accretion of few larger survivors, ultimately the present planets. 

Mars accreted very rapidly with a mean accretion lifetime of 1.9$^{+1.7}_{-0.8}$ million years \citep{dauphas2011hf,tang201460fe}. Similarly, the time scale of accretion and metal silicate differentiation of most of the NC family of iron meteorite parent bodies ranges from $\sim$$t_{CAI}$ to 2 Ma \citep{hilton2019genetics,hellmann2024hf}.Therefore, these bodies formed contemporaneously with and during the lifetime of PPD.  Given its Mars-like size, it is possible that a similar time sequence can be invoked for Mercury. However, questions remain: How long did Earth's accretion last, and what role did electromagnetic forces play in controlling this process?

The condensation of elements in the protoplanetary disk is controlled by two factors: temperature and gas fugacity, the latter of which controls the chemical behavior of an element (i.e. lithophile preferentially incorporated into oxides/silicates, siderophile in Fe-Ni alloys and chalcophile in sulfides) \citep{wasson1988compositions,lodders2003solar}. The sequence of mineral formation from a cooling protoplanetary disk, particularly for the abundant components (O, Fe, Mg, and Si), depends on the condensed fraction as a function of T and P, which is typically represented by the $T_{50}$ condensation temperature. For the inner solar system, high-temperature (or refractory) phases include an Fe-Ni alloy, olivine, anorthite, and both high- and low-Ca pyroxene.

\begin{figure*}[h]
\centering
\includegraphics[width=0.9\linewidth]{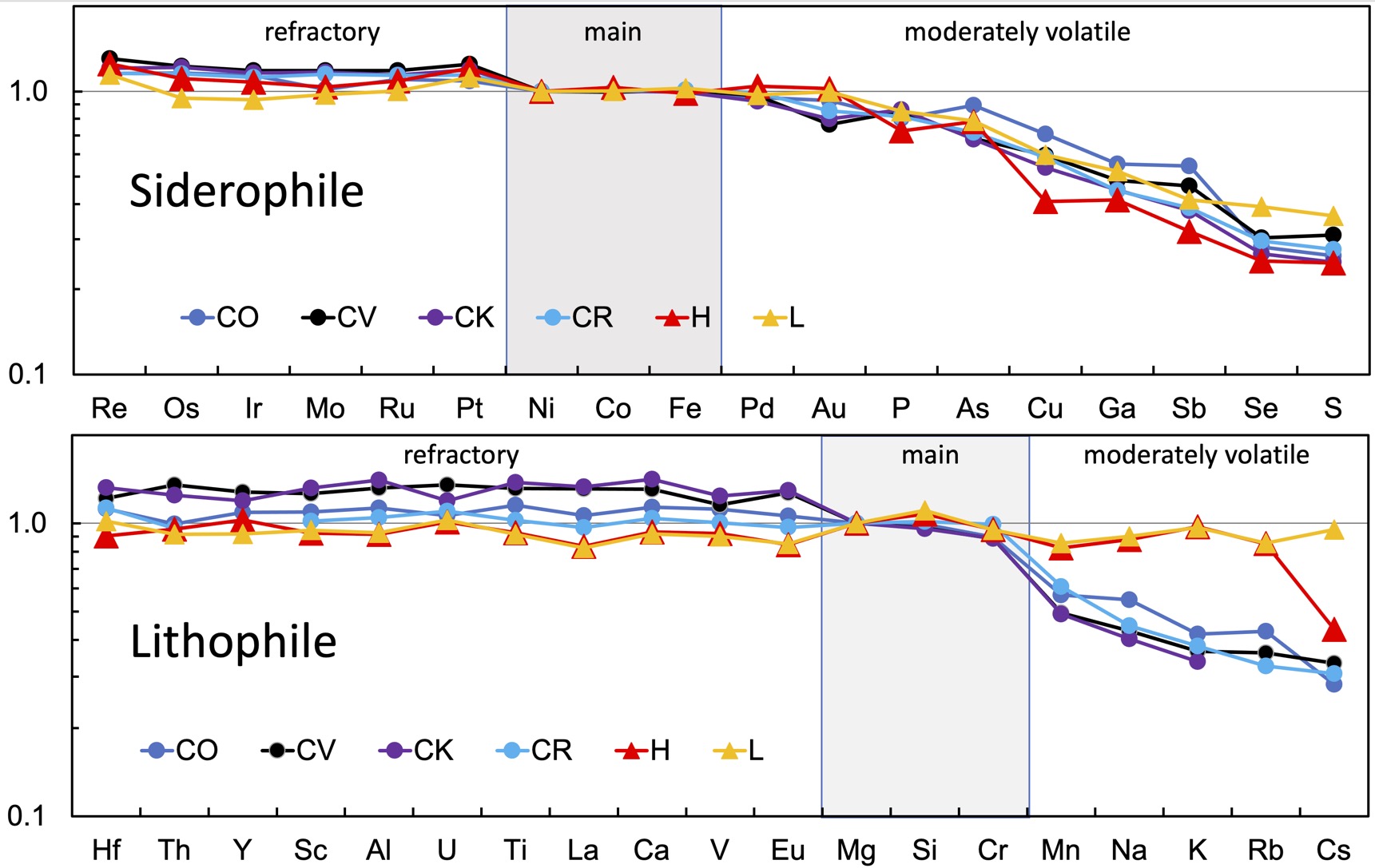}
\caption{Comparison of the CI normalized patterns of elements in carbonaceous (circle data points) and ordinary (triangle data points) chondrites. Siderophile and lithophile elements are normalized to Ni and Mg, respectively; these are two abundant main elements for such comparisons. Element order follows the $T_{50}$  condensation temperatures (range of values given just above x-axis) as reported in \cite{lodders2003solar} for refractory and main elements. Note the constancy of Ni-Co-Fe in all chondrites and their shared normalized pattern for the moderately volatile, siderophile elements (Au - S). The moderately volatile lithophile elements in contrast display markedly divergent behaviors.}
\label{fig:sid-lith}
\end{figure*}

The regularly used models for $T_{50}$ condensation temperatures \citep{wasson1985meteorites,lodders2003solar,wood2019condensation} agree for refractory and main elements (Figure \ref{fig:sid-lith}). There are differences between models for the condensation estimates for volatile elements \citep{wasson1985meteorites,lodders2003solar,wood2019condensation}. The \cite{wood2019condensation} revision of $T_{50}$ condensation temperatures were based on newly determined abundances for the halogens in CI chondrite \citep{clay2017halogens}. However, later it was shown that these halogen data were incorrect \citep{palme2022composition,lodders2023solar}. Consequently, I do not recommend using the \cite{wood2019condensation} $T_{50}$ condensation temperatures for volatile elements. An ongoing debate is centered around the appropriate $T_{50}$ condensation temperature for In, it has an estimated $T_{50}$ condensation temperature between 470 K \cite{wasson1985meteorites} and 536 K \cite{lodders2003solar}, or even a higher condensation temperature of 800 K \citep{righter2017distribution}.

Figure \ref{fig:sid-lith} shows the relative behavior of siderophile and lithophile elements in carbonaceous and ordinary chondrites. These chondrites have a range of Mg, Fe, Ni, S, and alkali element contents and display depletions in moderately volatile elements comparable to those observed in terrestrial planets. In contrast, the abundance patterns of the siderophile and lithophile elements for the enstatite chondrites are unlike those seen in any of the terrestrial planets (see \cite{braukmuller2025moderately} for a similar conclusion) and thus are not included in Figure \ref{fig:sid-lith}.

Refractory elements, for both siderophiles (and chalcophiles) and lithophiles, show flat unfractionated patterns (Figure \ref{fig:sid-lith}), which is a well-known attribute of these elements \citep{wasson1988compositions}. Moderately volatile elements show different behaviors for siderophiles and lithophiles. Although lithophile elements in ordinary chondrites have flat unfractionated patterns, carbonaceous chondrites have temperature-dependent depletions in moderately volatile elements. Siderophile elements, in both carbonaceous and ordinary chondrites, display a more coherent depleted pattern in the absolute and relative abundances of moderately volatile elements.
 
The Fe-Ni alloy, the host for Co, Au, As, Ga and others, acts as a solid solution host phase for these minor and trace elements and so once formed, Fe metal takes in other siderophilic elements, while still maintaining half mass temperature condition with half of the element in the gas and half in the host condensate \citep{lodders2003solar}.  However, it has also been suggested that compositional contrasts between lithophiles and siderophiles (and refractories and main elements vs moderately volatiles) might reflect differences in settling and accretion efficiencies in the midplane of the protoplanetary disk \citep{wasson1988compositions}. More recently, we highlighted that fractionation in the protoplanetary disk might also be influenced by the higher density, magnetic susceptibility of Fe, Ni alloys, and contrasting elastic properties of metals (vs silicates) \citep{larimer1970chemical}, which in turn influence the compositions of chondrites and terrestrial planets \citep{mcdonough2021tp}.

Carbonaceous chondrites contain abundant CAIs, however, these inclusions are rare in enstatite and ordinary chondrites (NC) \citep{dunham2023calcium}. The rarity of CAI in NC chondrites might be consistent with the formation of a gap in the protoplanetary disk that trapped CAIs in the outer solar system \citep{dunham2023calcium}, and / or the incorporation of CAIs into early formed inner solar system planets, leaving rare levels of inclusions for later incorporation into NC chondrites. Relative to the NC chondrites, carbonaceous chondrites are enriched in refractory lithophile elements, while NC and CC chondrites have comparable levels of refractory siderophile elements (Figure \ref{fig:sid-lith}). This latter observation is consistent with a common carrier for siderophilic and chalcophilic elements (that is, the Fe, Ni alloy) \citep{wasson1988compositions}. Compounds of the Fe, Ni alloy began condensing with refractory elements and continued to do so over a wide range of $T_{50}$ condensation temperatures up to the appearance condensation temperature of FeS at 704 K \citep{larimer1967chemical,lodders2003solar}. Thus, a broad range of siderophile elements (e.g. P, Cu, Ga, Ge) are likely to condense into a Fe, Ni alloy in the protoplanetary disk. Lithophile elements, particularly the moderately volatiles, are not restricted to a common phase but appear to be carried less by some chondrules and more so by volatile matrix materials \citep{braukmuller2018chemical,bland2005volatile,van2019unifying}. For the Earth, the proportion of volatile-rich matrix material was considered to be $\leq$20\% \citep{yoshizaki2021earth}. 

Differentiation of metal and silicate in early-formed planetary bodies was initially driven by the presence of $^{26}$Al and $^{60}$Fe; smaller bodies were challenged to cool and crystallize because they were robustly heated by nuclear and collisional processes \citep{mcdonough2020radiogenic} (see also Sections \ref{nu} and \ref{thermal}). The physical properties of these molten or crystallized additions affected their re-equilibration and re-incorporation.

\section{Pitfalls with assuming a solar composition for bulk planet compositions}

Although differences in the proportion of the eight elements that make up the rocky planets and chondrites exist, a simple compositional model for a planet can be constructed based on the composition of the solar photosphere and CI chondrite, the most primitive chondrite composition. 

\begin{wrapfigure}{r}{0.55\textwidth}
\centering
\includegraphics[width=.90\linewidth]{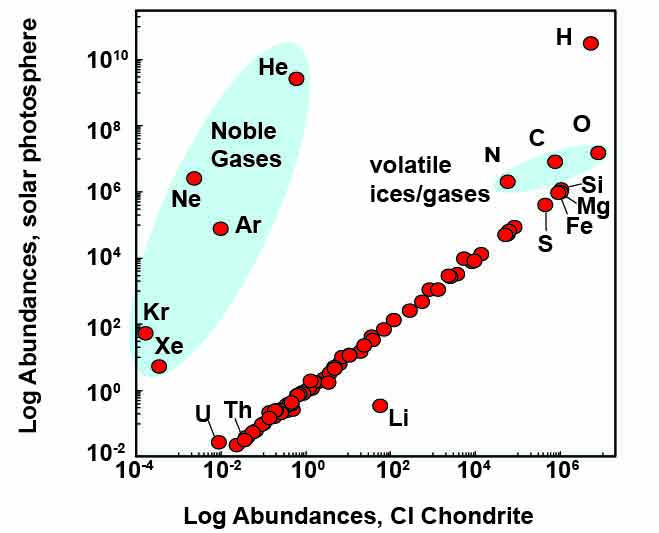}
\vspace{-2mm}
\caption{Log abundances of CI chondrite versus the solar photosphere based on a million atoms of Si in each. Data from \cite{magg2022observational,lodders2021relative}.}
\label{fig:photo}
\end{wrapfigure}

\begin{wrapfigure}{R}{0.5\textwidth}
\centering
\includegraphics[width=.90\linewidth]{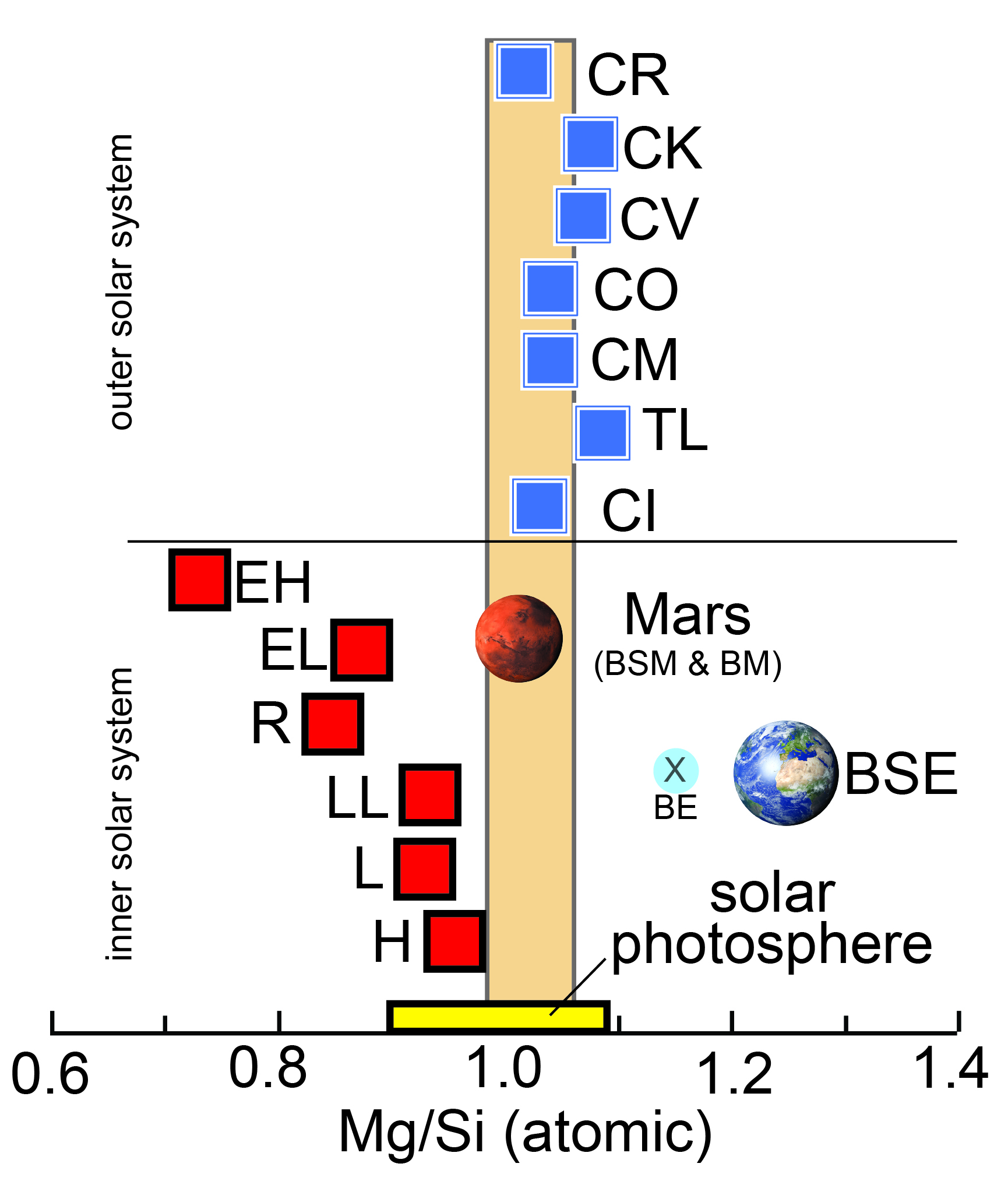}
\caption{Mg/Si variation in NC (red) and CC (blue) chondrites, Mars (bulk silicate Mars (BSM) and bulk Mars (BM)), Earth (bulk silicate Earth (BSE) and bulk Earth (BE)) and the solar photosphere. Data sources: chondrites \citep{wasson1988compositions}, solar photosphere \citep{magg2022observational}, Mars \citep{yoshizaki2020composition}, BSE and bulk Earth \citep{mcdonough1995,mcdonough2021tp}.}
\vspace{-5mm}
\label{fig:mgsi}
\end{wrapfigure}

\cite{goldschmidt1938mengenverhaltnisse} was the first person to observe the compositional match between the solar photosphere and CI chondrite. Today it is recognized that this 1:1 correlation extends over six orders of magnitude for all but the highly volatile elements and the noble gases, when normalized to a million Si atoms (Figures \ref{fig:photo}). This compositional match provides a powerful observation, as the Sun is the primary mass of the solar system (see Table \ref{table:T1}), and the rocky planets collectively contributed only a few parts per million in absolute mass. These observations have led many, particularly those in the mineral physics and geodynamic communities, to make the mistaken assumption that rocky planets and the Sun possess the same mass ratio of elements (e.g., Mg/Si, Fe/Mg) (Figures \ref{fig:mgsi}). Here we show that this ad hoc assumption is arbitrary and without justification.

\begin{table}[b!]
\centering
\caption{Compositional comparisons of CI and Ordinary Chondrite (OC) planets and the Earth}
\begin{tabular}{c|ccc|cc|cc}
\\[-2ex]
\hline
  &   & $^*$Bulk   & Bulk Silicate   & Bulk    & Bulk Silicate  & Bulk  & Bulk Silicate  \\
 wt \% & \textbf{CI} & ``CI planet" &  ``CI planet"  & ``OC planet"    &  ``OC planet" & Earth & Earth \\
      \hline
O     & 46.0 & 30.0   & 45.0     & 35.7  & 44.0  & 30.0  & 44.0  \\
Na    & 0.51 & 0.77   & 0.3     & 0.6   & 0.3   & 0.18  & 0.27  \\
Mg    & 9.65 & 14.5   & 21.7     & 14.0  & 21.0  & 15.4  & 22.8  \\
Al    & 0.86 & 1.29   & 1.94     & 1.1   & 1.70  & 1.59  & 2.22  \\
Si    & 10.7 & 16.0   & 24.0     & 16.9  & 25.4  & 15.3  & 21.0  \\
S     & 5.40 & 8.10   & 0.02     & 2.0   & 0.02  & 1.10  & 0.02  \\
Ca    & 0.92 & 1.38   & 2.07     & 1.3   & 1.88  & 1.71  & 2.34  \\
Fe    & 18.2 & 27.3   & 6.26     & 27.5  & 6.50  & 31.9  & 6.26  \\
Ni    & 1.07 & 1.61   & 0.20     & 1.6   & 0.20  & 1.83  & 0.20  \\
\hline
total & 93.3 & 100.9  & 101.4     & 100.7 & 100.9 & 98.9  & 99.1   
\end{tabular}
\par\medskip \raggedright
$^*$Bulk ``CI planet" = CI $\times$ 1.5, less correction for O loss due to dehydration and decarbonization. \\
Bulk Silicate ``planet" = Bulk planet $\times$ 1.5, less Fe, Ni, and S to the core. 
Data for CI and OC (H type) from \cite{wasson1988compositions} and Earth from this study and \cite{mcdonough1995}. 
\label{table:CI-E}
\end{table}

An often proposed simple model for building a rocky planet uses a CI chondrite composition (Table \ref{table:CI-E}). CI chondrites have high water and CO$_2$ contents relative to a rocky planet. Consequently, the devolatilization of CI can lead to a bulk enrichment factor of about 1.5 in the abundance of elements in the planet. The absolute value of this enrichment factor is not fixed, but requires an understanding of the Earth's content of water and CO$_2$. Likewise, core formation results in the bulk silicate fraction being enriched in lithophile elements by another factor of about 1.5 for a 1/3 metal fraction. Together, this results in an enrichment factor of 2.2 in the silicate shell for the refractory lithophile elements. Fe, Ni and S in the silicate shell are reduced due to core subtraction as well as Na due to its volatility. Here we set the amount of Fe and Ni to be comparable to that of the Earth's mantle, although this is not required.  Modeling the abundances of the moderately volatile elements (e.g., Na and S) in the planet depends on data (i.e., meteorites and orbital surveys) used to establish a planetary volatility curve (e.g., Na, K, Rb). The planetary inheritance of sulfur can differ by a factor of 4 depending on the estimate of volatile depletion for the planet. For example, the estimated S content of the bulk Earth is about 1 wt\% \cite{fischer2025earth}, while estimates of Mars range up to $\sim$4 wt\% \citep{yoshizaki2020composition}.

A model can also be constructed using an H type Ordinary Chondrites (OC, Table \ref{table:CI-E}).  This chondrite type is rich in Fe and poor in water and CO$_2$, so the first step of devolatilization is not required. Again there is a 1.5 times enrichment factor for lithophile elements in the bulk silicate due to extraction of a 1/3 mass fraction for the core. Both chondrite models produce reasonable compositions, however, the latter two model planets have much lower Mg/Si, Mg/O and Fe/O values than that seen in the BSE, which would lead to markedly different mineralogical proportions in Earth's lower mantle. In addition, these two chondritic models for planets have lower radiogenic heat production due to lower amounts of Th and U (refractory lithophile elements), which contribute 80\% of the current heat production \citep{mcdonough2020radiogenic}).

Compared to a CI chondrite model, the bulk silicate Earth (Table \ref{table:CI-E}) is estimated to have an enrichment factor of $\sim$2.6 for the refractory lithophile elements in the BSE or $\sim$1.9 for the refractory elements (e.g., Mo, W, Ca, and Al) in the bulk planet \citep{mcdonough1995}. Both Earth and Mars have similar levels of enrichment of refractory elements of $\sim$1.9, despite differences in Fe and Ni contents (i.e., core fraction) \citep{yoshizaki2021earth}. This enrichment in refractory elements, a possible characteristic of rocky planets, may be due to early planetary growth that incorporated more refractory than main elements in the condensing phases. 

Although this simple compositional modeling produces reasonable estimates for the rocky planets, the results are not comparable to established observations for Earth and Mars \citep{mcdonough2021tp}. Whereas CI chondrite has an O/Fe value of $\sim$8.8 (atomic ratios), the Earth's value is $\sim$3.3 \citep{mcdonough2021tp} and Mars is $\sim$5.4 \citep{yoshizaki2020composition}. The oxygen isotopic composition of the Sun relative to the planets and chondrites is starkly different \citep{mckeegan2011oxygen}. Recent measurement of the solar wind by the Genesis experiment \citep{burnett2025solar} determined a solar system Fe/Mg, 0.79 $\pm$ 0.05 that, within uncertainties, agrees with the CI chondritic (0.82) and spectroscopic photospheric (0.82) values, but not with that of the bulk Earth (0.90) and Mars (0.67) \citep{mcdonough2021tp}. 

If NC chondrites are considered as representative of the inner solar system, their Mg/Si value is neither CI chondritic, photospheric, nor Earth like; the distinctive Mg/Si values of inner solar system chondrites are consistent with these materials not being isochemical precipitates from the protoplanetary disk, but were late condensates that followed a crystallization sequence as documented in chondrites \citep{grossman1972condensation}. The silica-rich nature of ordinary and enstatite chondrites relative to CI chondrites is consistent with the former having more modal pyroxene, which reflects cooler conditions of condensation from the PPD. 

Two common, but incorrect statements are often made: ``the Earth has a chondritic Mg/Si ratio", or ``we assume the Earth's Mg/Si is equal to that of the solar photosphere". The former sentence is incorrect, as chondrites do not have a constant Mg/Si ratio; it varies by some 30\% (Figure \ref{fig:mgsi}). The outer solar system has a near-constant Mg/Si value, which matches that of the Sun's photosphere, while the inner solar system shows marked variations in Mg/Si. The second incorrect statement (...assume the Earth's Mg/Si...) assumes an isochemical transition for planets from the protoplanetary disk, which is at odds with the abundant evidence for olivine followed by low Ca pyroxene crystallization as the protoplanetary disk cools. The proportion of olivine (Mg$_2$SiO$_4$) to low Ca pyroxene (MgSiO$_3$) in rocks and planets reflects its bulk Mg/Si value. The distinctly lower Mg/Si values for inner solar system chondrites reflect their higher pyroxene/olivine proportion, a likely consequence of later cooler condensation in an evolving protoplanetary disk. All NC chondrites (that is, inner solar system chondrites) have significantly lower Mg/Si values than CI and other CC chondrites, while that for Mars is similar to the photosphere's value and Earth's is higher (Figure \ref{fig:mgsi}).

\section{Composition of the Earth and bulk silicate Earth}

The composition of rocky planets like Earth can be determined by combined chemical and isotopic analyses of mantle-derived samples and comparisons with chondrites, the remnants of later accretion. More than 90\% (atomic and/or weight) of these cosmic rocks are composed of O, Fe, Si, and Mg, with Fe being the most variable component; there is more than a factor of three differences in Fe contents between chondrites \citep{mcdonough2021tp}. Adding Ca, Al, Ni and S accounts for 99 wt\% of the mass of the rocky planets (Figure \ref{fig:8elem}), with Ca/Al and Fe/Ni ratios fixed (1.07 $\pm$ 0.04 and 17.4 $\pm$ 0.5, respectively) using constant chondritic ratios. Using chemical correlations in peridotites, \cite{mcdonough1995} established chondritic proportions of the refractory lithophile elements in the mantle and determined their absolute abundances (i.e., the amount of Ca, Al, and Ti). They also noted similar Si contents for basalts and residual peridotites, consistent with a Si bulk distribution coefficient of $\sim$1 and a mantle composition of $\sim$45.0 wt\% SiO$_2$. In addition, the FeO contents of primitive basalts are approximately 15\% higher than those of the residual peridotites. From this and knowing the Mg number of the mantle (atomic ratio of Mg/Mg + Fe = 0.89), its Ni content ($\sim$1960 ppmw (parts per million by weight)) and the mass and density of the planet and/or the core, one can establish the bulk composition of a rocky planet.

Here, I re-evaluate the compositional model for the bulk silicate Earth (aka pyrolite, primitive mantle) presented in \cite{mcdonough1995}. Since the publication of that model, there have been considerable advances in geophysics and geochemistry of the Earth and in the analysis and understanding of chondritic and nonchondritic meteorites, particularly the identification and understanding of NC and CC meteorites and the establishment of the isotopic dichotomy \citep{warren2011stable,bermingham2020nc,kruijer2020great}. Importantly, starting in 2005 \citep{araki2005experimental} with the first measurement of the Earth's geoneutrino flux, particle physicists gave us the unparalleled opportunity to measure the whole Earth's U and Th content, the only method for observing a global geochemical signal of the planet in real time.

\subsection{Earth's Nd isotopic composition}

\begin{wrapfigure}{r}{0.6\textwidth}
\centering
\includegraphics[width=0.95\linewidth]{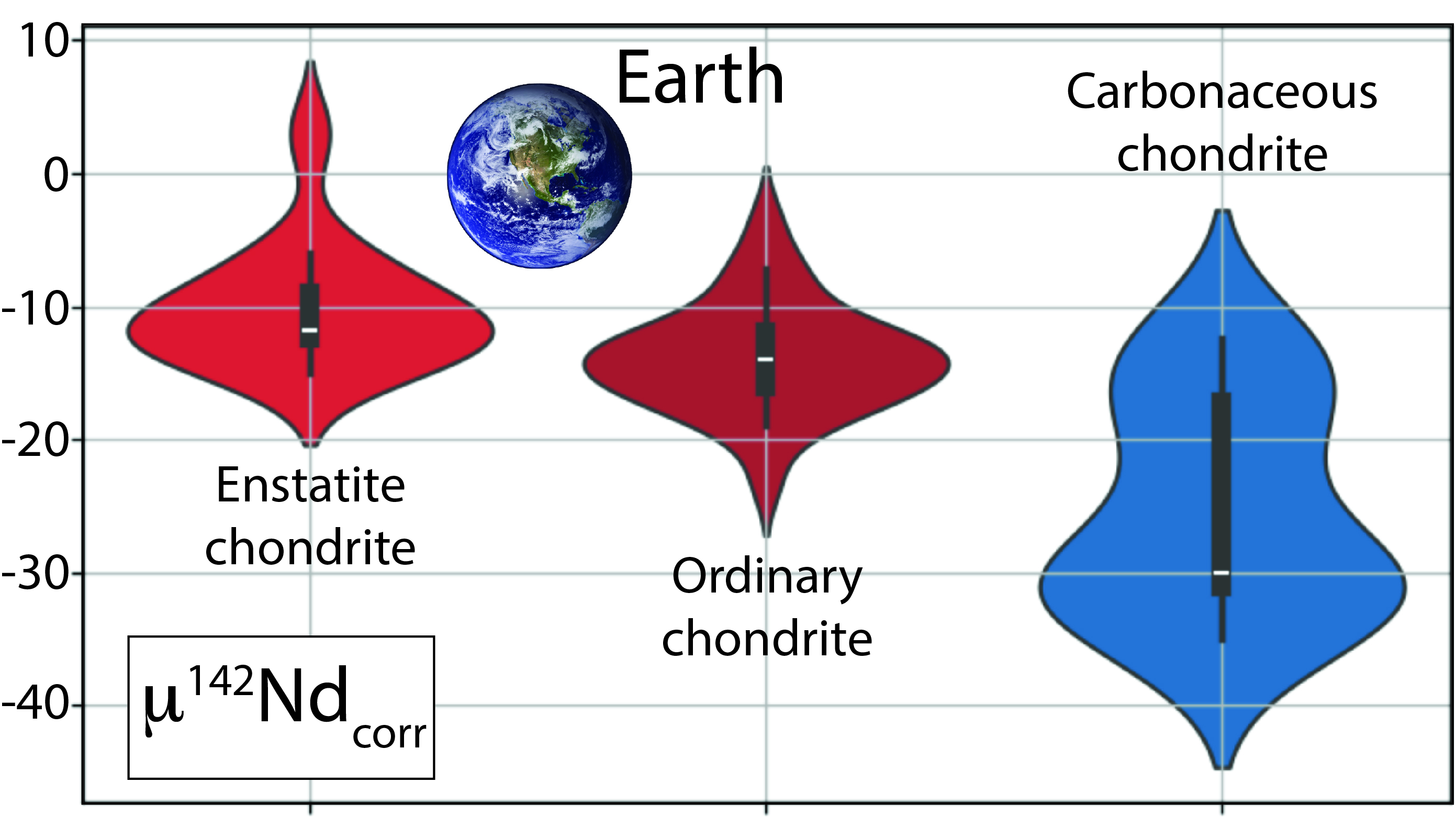}
\caption{Violin plot for $\mu^{142}Nd_{corr.}$ in chondrites and Earth (see text for details). Data for NC (Enstatite and Ordinary chondrites; inner solar system) and CC (Carbonaceous chondrites; outer solar system) meteorites are from \cite{frossard2022earth}. }
\label{fig:142}
\end{wrapfigure}

In their initial report on the short-lived, extinct $^{142}$Nd isotope system ($^{146}$Sm $\rightarrow$ $^{142}$Nd + $\alpha$ + $Q$; t$_{1/2}$ = 103 Myrs) in NC and CC meteorites, \cite{boyet2005142nd} found that the BSE's $\mu^{142}Nd_{corr.}$ value (i.e. 0) is 20 ppm higher (where $\mu^{142}Nd_{corr.}$ represents ppm deviations from Earth's value as a terrestrial standard and corrected (corr.) for decay of $^{146}$Sm) than what they found in chondritic meteorites. The implication being that one or more dynamical processes  occurred in the early Earth which caused a shift in the accessible Earth's Nd budget. Thus, to account for a balanced isotopic composition of Nd in the bulk Earth, they invoked a missing or hidden component that has the complementary isotopic composition. Since then, other studies \citep{frossard2022earth,johnston2022nd} have designated chondrite groups with narrow definitions for $\mu^{142}$Nd to predict the composition of the Earth. Figure \ref{fig:142} shows the isotopic variation of $\mu^{142}$Nd for chondrites and the Earth. On average, the inner and outer portions of the solar systems are compositionally offset, with the Earth's composition plotting at the end of the compositional spectrum. Representing an end-member in compositional space, the bulk Earth's $\mu^{142}$Nd isotopic composition of 0 is consistent with other measured nucleosynthetic anomalies (i.e., Ti, Cr, Zr, Mo, and Ru) where the Earth's value is at or towards the end of the range of compositions \citep{burkhardt2021terrestrial,bermingham2017ruthenium,fischer2017ruthenium,render2022solar,bermingham2024non}. Collectively, these observations can be interpreted as reflective of the bulk Earth's $\mu^{142}$Nd isotopic composition without the need to invoke any lost or hidden component. 

Alternative models \citep{boyet2005142nd,frossard2022earth,johnston2022nd} for interpreting the Earth's $\mu^{142}$Nd value relative to NC chondrites suggest that the BSE's value is not zero, but slightly negative (i.e., somewhere between -5 and -15 ppm), with the result of $\epsilon_{^{143}Nd/^{144}Nd}$ being about +3 for the accessible Earth (where $\epsilon_{^{143}Nd/^{144}Nd}$ represents parts per 10$^4$ deviations from Earth's value as a terrestrial standard, and accessible Earth represents crust and mantle analyzable samples).  This latter Nd isotope system ($^{147}$Sm $\rightarrow$ $^{143}$Nd + $\alpha$ + $Q$; t$_{1/2}$ = 106 Gyrs)  is the long-lived system that has been used for decades to characterize the Depleted Mantle (generally +5 to +12 $\epsilon_{^{143}Nd}$), the source of mid-ocean-ridge basalts and Enriched Mantle, the source of ocean island basalts (generally +8 to -3 $\epsilon_{^{143}Nd}$). The term ``accessible Earth" refers to what is accessible as opposed to a fraction of the silicate Earth that was either lost to space via collisional erosion or transported and isolated to the base of the mantle. A consequence of collisional erosion, if it occurred, is a significant reduction in the BSE of the heat producing elements (HPE: K, Th and U) and other highly incompatible elements (D$_{sol/liq}$ $\ll$ 1) that would have been concentrated in the first crust to crystallize on Earth \citep{boyet2005142nd,frossard2022earth}.  Depending on the amount of offset needed in the $\mu^{142}$Nd isotopic composition, there is an even greater number of highly incompatible elements, including HPE, that have been lost \citep{oneill2008collisional,frossard2022earth}. Recently, \cite{hofmann2022size} have argued for the sequestration of an early enriched reservoir (EER) in the silicate Earth (but also allow it to be lost to space). This model is similar to, but different, from the above and related models. Likewise, the creation of this reservoir in the silicate Earth leads to the present-day BSE having an $\epsilon_{^{143}Nd}$ of $\sim$+3. On the other hand, a compelling Nd-Ce-Hf isotopic observation has been made by \cite{willig2019earth} and \cite{stracke2025geochemical}, where they argue that the coincidence of this triple radiogenic isotope system at the 0 intercept is consistent with Earth's $\mu^{142}$Nd value of 0. Collectively, these observations provide competitive interpretations but are insufficient to argue against any one of these models. Future geoneutrino studies (see Section \ref{nu}) will be able to set the current budget of Th and U in the planet and limit the range of acceptable models.

\subsection{The form of mantle convection}
Modern seismic tomographic images reveal the planform of mantle convection. Images of subducting slabs of oceanic lithosphere that stagnate above, penetrate through, and are trapped below the 660 km seismic discontinuity capture the convective dynamics of Earth in action and the final fate of plate tectonics \citep{fukao2013subducted}. These images document mass exchange between the upper and lower mantle and reveal that the ultimate fate for some slabs is deposition into the deepest parts of the lower mantle. In addition to documenting a whole mantle mode of convection, these images document that there is no conductive boundary layer between the upper and lower mantle, which would be required if the mantle is compositionally layered. Moreover, given the mantle's two recognized conductive layers at the top and bottom of the mantle (that is, the lithophere and the D`` layer), one might anticipate a globally encircling 100 - 200 km thick conductive layer in the Transition Zone (that is, between the 410 km and 660 km seismic discontinuities) isolating the convection modes of the upper and lower mantle. This structure has never been globally identified. \textit{Thus, given whole mantle convection, compositional models of the mantle can be treated as relatively homogeneous at the level of the major elements.} Although not compositionally layered, the mantle contains structures that appear to be compositionally distinct and range in size from small to continental and larger \citep{garnero2016continent,cottaar2016morphology,ballmer2017persistence,talavera2025global}. For example, seismic images capture mantle plumes that rise from the base of the mantle up to about 1000 km depth and spread out as they penetrate the seismic discontinuities at 660 and 410 km depth (for example, \citep{french2015broad}).

Some would argue that the above perspective of whole mantle convection may be an incomplete picture. The deepest 500 km or so of the mantle may have compositional domains that deviate from an average pyrolitic composition. These models are based on what can be deduced from extrapolated seismological and mineral physics data to modal mineralogy. A pyrolite mantle would have approximately 7 to 8 mode\% davemaoite (that is, Ca-perovskite), $\sim$ 17 mode\% ferropericlase and $\sim$ 75\% mode\% bridgmanite, while a BSE model derived from a chondritic composition has little or no ferropericlase, possibly some stishovite (SiO$_2$ polymorph), davemaoite, with the proportions depending on the amount of CaO, and the remainder being mostly bridgmanite. 

Proposals for compositional layering in the mantle range from small to large scale. Some envisage the lower 400 km of the mantle as having an enrichment in silica possibly due to the remains of an ancient basal magma ocean \citep{cobden2024full}. More recently, it was suggested that LLSVPs (Large Low Shear Velocity Province) originated early in the history of the Earth and are enriched with bridgmanite \citep{talavera2025global}, while \cite{panton2025unique} cannot rule out the possibility that primordial chemically distinct material is present at the base of LLVPs. More recently \cite{murakami2024composition} have stated that mineral physics-informed profiles of seismic properties, based on a lower mantle of bridgmanite and ferropericlase, are consistent with
Mg/Si 0.9–1.0 when compared to radial seismic reference models. On a larger scale, primordial structures, rich in bridgmenite (and depleted in ferropericlase) and up to 10\% to 15\% of the mass of the mantle mass, have been invoked to justify solar-chondritic Mg/Si ratios \citep{ballmer2017persistence}.  

Overall, the conclusions reached in the above studies challenge the pyrolite model presented here. However, the strength of these conclusions depends on the combined uncertainties of the EOS (equation of state) data, the thermal profile assumed for the deep mantle, and the seismic velocities under which the calculations were carried out. Currently, the present state of this modeling is not mature enough to reach conclusions that can discriminate between competing compositional models.
 
\subsection{Crust-mantle melting to determine the BSE composition}

\begin{wrapfigure}{r}{0.6\textwidth}
\centering
\includegraphics[width=.95\linewidth]{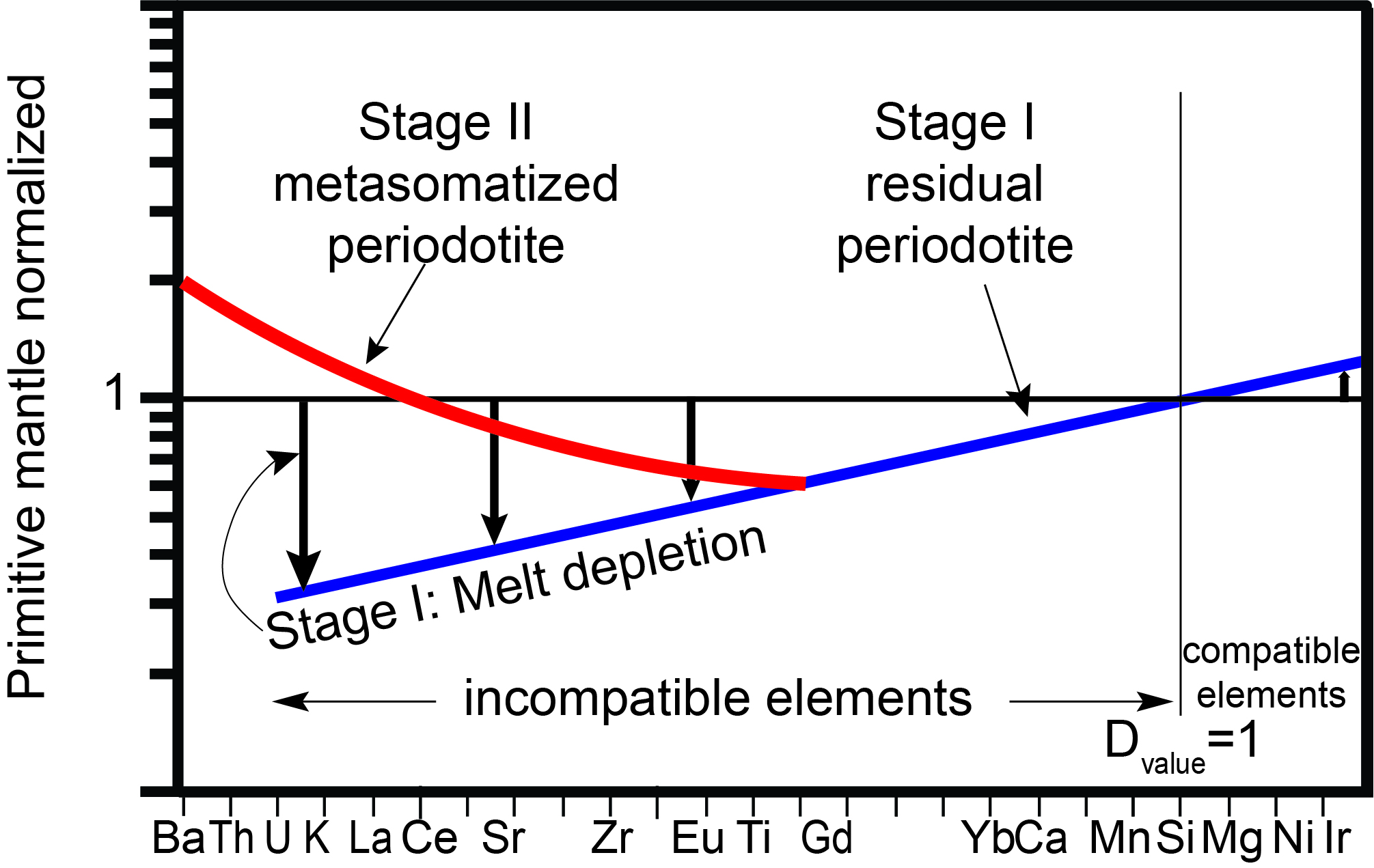}
\caption{An idealized mantle-normalized diagram illustrating the typical two stage evolution of residual mantle peridotites. First, melt production results in the depletion of incompatible elements (i.e., D$_{values}<$1, concentration in the residue is less than in the melt)and slight enrichment of compatible elements (blue trend). Second, there is a re-enrichment of the most highly incompatible elements (red trend). Under most conditions Si has a D$_{value}$=1, concentration in the melt and residue are equal and it is a pivot point on the diagram. Timing of the first and second stage events can be nearly simultaneous or not.}
\label{fig:peridotite}
\end{wrapfigure}

Partial melting of the mantle produces residual peridotites with mild enrichments (5\% to 50\%) in compatible elements and slight depletions (less than a factor of three) in mildly incompatible elements \citep{sun1989chemical,mcdonough1994chemical}. Mantle rocks with these characteristics (generally those having a MgO content $<$41wt\%) can be used to constrain the composition of the bulk silicate Earth \citep{mcdonough1995}. The composition of moderately incompatible elements (for example, Ti, Eu, Gd) and elements that are more incompatible (K, U, and Th) are less reliably established by these methods due to secondary mantle processes (that is, metasomatism) that overprint the compositions of residual peridotites (Figure \ref{fig:peridotite}). In this case, studies of basalts and rocks from continents can be used to obtain insight into the absolute abundances of these more incompatible elements.

Peridotite samples were used to identify the characteristics of the primitive mantle. To this end, \cite{mcdonough1995} established two criteria for sample selection: 1) MgO content $\leq$ 40.5 wt \% and 2) no evidence of metasomatic enrichment (chondrite-normalized [La / Yb]$_N <$ 2). In addition, analysis of the data were restricted to elements that are only slightly incompatible during partial melting, as these are the least altered by later metasomatic reenrichment \citep{mcdonough1994chemical} (that is, avoiding elements affected by the overprinting shown by the red pattern, Figure \ref{fig:peridotite}). It appears that all residual peridotites, massifs and xenoliths, experience melt extraction and limited reenrichment by a metasomatic fluid, some of which can be associated with the original melting event. \cite{lyubetskaya2007chemical} using this exact same dataset as \cite{mcdonough1995}, but with different selection criteria (MgO content $\leq$ 43 wt\% and [La / Yb]$_N <$ 2) determined a slightly different composition of the BSE, although it was within uncertainties equal to that determined by \cite{mcdonough1995}. 

However, their statistical method is more sensitive to outlier data and their Table 1 analyzes show that they used more incompatible elements, the middle and light REE, to perform regression analyzes, a practice avoided by \cite{mcdonough1995} as these elements were influenced by metasomatic enrichment processes (Figure \ref{fig:peridotite}). The more restrictive approach used by \cite{mcdonough1995} produces higher uncertainties, but more accurate results. 

Table \ref{tab:BSE} represents an update to the original compositional model developed by \cite{mcdonough1995}. The bulk silicate Earth was determined to have refractory lithophile elements at 2.65 times that of CI carbonaceous chondrites. Updated concentrations and uncertainties for the following elements, which were established by dedicated studies that explored geochemical proxies distinct from those first leveraged by \cite{mcdonough1995}, include: Li and B \citep{marschall2017boron,walowski2019investigating}, P (using P/Pr) \citep{jenner2012analysis}, W \citep{arevalo2008tungsten}, Sn \citep{witt2009geochemistry}, Cd \citep{yi2000cadmium,witt2009geochemistry}, Ag \citep{wang2015abundances}, Se, and Te \citep{wang2013ratios}, halogens \citep{kendrick2017seawater}, HSE (highly siderophile elements) \citep{becker2006highly,fischer2011rhodium}, and U was set at 0.265 times Th \citep{wipperfurth2018earth}.  

In the last 30+ years, much has been written about the abundance of Nb in BSE, and whether or not BSE has a chondritic Nb/Ta value \citep{mcdonough1991partial,rudnick2000rutile,jochum2000niobium,pfander2007nb,hofmann2022size}. Subgroup V of the periodic table, the vanadium group elements, is composed of V, Nb and Ta. In the 1980s and 1990s, estimating V abundance in the BSE was straightforward \citep{mcdonough1995}, as there was enough data of sufficient quality to show that it had $\sim$82 ppmw V and that this refractory element was not entirely lithophile, as only about 60\% of its total planetary budget was in the BSE. However, the same data state was not true for Nb and Ta at that time, as the data were few for chondrites, peridotites, komatiites, and basalts, and most were not of high enough quality. Today, the situation has improved greatly, particularly for chondrites \citep{munker2003evolution,barrat2023trace} and basalts \citep{hofmann2022size}. Enstatite chondrites have a Nb/Ta of 19.7 $\pm$ 6.0 \citep{barrat2023trace}, comparable to carbonaceous chondrites 18.9 $\pm$ 1.9 \citep{munker2003evolution}.  In comparison, the accessible Earth appears to have a Nb/Ta of 15.5 based on data for basalts and crustal rocks \citep{hofmann2022size}. Therefore, I have decreased the abundance of Nb in BSE by $\sim$ 20\% to 0.605 ppmw (Table \ref{tab:BSE}, cf. \citep{mcdonough1995}). The elements of subgroup V appear to show a systematic change in their lithophile to siderophile behavior downward in this column, with elements becoming more lithophile with higher Z. Relative to the bulk Earth, it appears that $\geq$40\% of V and $\sim$20\% of Nb is hosted in the Earth's core. This conclusion is consistent with experimental studies \citep{wade2001earth,huang2020niobium}.

The compositional model approached used here and by \cite{mcdonough1995} takes into account any and all differentiation events experienced by the BSE. This would include melt-residue processes that might have been involved in the production of an early enriched (and depleted) reservoirs \citep{boyet2005142nd,hofmann2022size}. Therefore, this calculated BSE composition recovers the starting condition for the Earth.

\begin{table}[H]
\caption{Compositional model of bulk silicate Earth (BSE), updated from McDonough and Sun (1995)}
\label{tab:BSE}
\centering
\begin{tabular}{lc|cc|cc||lc|cc|cc}
\\[-2ex]
\hline
  & CI        & BSE    & BSE          &  &    &  & CI        & BSE  & BSE      &  &     \\
  & chond. & ppmw    & $\mu$moles/g & $\pm$ & $\Delta$   &  & chond. & ppbw  & nmoles/g & $\pm$ & $\Delta$    \\
  \hline
H     & 18530     & 100    & 99          & F3    & -400  & Rh   & 0.131     & 1.2  & 0.01     & 15    & 25   \\
Li    & 1.48      & 1.6    & 0.23         & 20    & 0   & Pd   & 0.563     & 7.1  & 0.07     & 15    & 45   \\
Be    & 0.0235    & 0.062  & 0.0069       & 10    & -9  & Ag   & 0.201     & 9    & 0.08     & 35    & 11   \\
B     & 0.81      & 0.30   & 0.028        & 30    & 0   & Cd   & 0.678     & 18   & 0.16     & 30    & -122 \\
C     & 34575     & 120    & 10           & F3    & 0   & In   & 0.080     & 11   & 0.10     & 40    & 0    \\
N     & 2725      & 2      & 0.14         & F2    & 0   & Sn   & 1.635     & 91   & 0.77     & 30    & -43  \\
O \%  & 46.1      & 44.0   & 27500        & 10    & 0   & Sb   & 0.144     & 5.5  & 0.045    & 50    & 0    \\
F     & 60.7      & 17     & 0.89         & 35    & -47 & Te   & 2.322     & 12   & 0.094    & 20    & 0    \\
Na    & 4926      & 2670   & 116          & 15    & 0   & I    & 0.520     & 7    & 0.055    & 60    & -43  \\
Mg \% & 9.51      & 22.8   & 9380         & 10    & 0   & Cs   & 0.187     & 21   & 0.16     & 40    & 0    \\
Al \% & 0.835     & 2.21   & 820          & 10    & -6  & Ba   & 2.411     & 6390 & 47       & 10    & -3   \\
Si \% & 10.70     & 21.0   & 7480         & 10    & 0   & La   & 0.2385    & 632  & 4.6      & 10    & -3   \\
P     & 1015      & 90     & 2.9          & 15    & 0   & Ce   & 0.6155    & 1631 & 12       & 10    & -3   \\
S     & 53920     & 250    & 7.8          & 20    & 0   & Pr   & 0.0921    & 244  & 1.7      & 10    & -4   \\
Cl    & 695       & 26     & 0.73         & 35    & 35  & Nd   & 0.4707    & 1247 & 8.6      & 10    & 0    \\
K     & 552       & 260    & 6.6          & 20    & 8   & Sm   & 0.1498    & 397  & 2.6      & 10    & -2   \\
Ca \% & 0.903     & 2.39   & 597          & 10    & -6  & Eu   & 0.0575    & 152  & 1.0      & 10    & -1   \\
Sc    & 5.83      & 15.5   & 0.34         & 10    & -5  & Gd   & 0.2016    & 534  & 3.4      & 10    & -2   \\
Ti    & 445       & 1180   & 25           & 10    & -2  & Tb   & 0.0367    & 97   & 0.61     & 10    & -2   \\
V     & 54.1      & 82     & 1.6          & 10    & 0   & Dy   & 0.2563    & 679  & 4.2      & 10    & 1    \\
Cr    & 2619      & 2625   & 50           & 15    & 0   & Ho   & 0.0559    & 148  & 0.90     & 10    & -1   \\
Mn    & 1886      & 1045   & 19           & 10    & 0   & Er   & 0.1635    & 433  & 2.6      & 10    & -1   \\
Fe \% & 18.49     & 6.26   & 1120         & 10    & 0   & Tm   & 0.0259    & 69   & 0.41     & 10    & 1    \\
Co    & 508       & 105    & 1.8          & 10    & 0   & Yb   & 0.1650    & 437  & 2.5      & 10    & -1   \\
Ni    & 10961     & 1960   & 33           & 10    & 0   & Lu   & 0.0248    & 66   & 0.38     & 10    & -3   \\
Cu    & 129       & 30     & 0.47         & 15    & 0   & Hf   & 0.1062    & 281  & 1.6      & 10    & -1   \\
Zn    & 314       & 55     & 0.84         & 10    & 0   & Ta   & 0.0148    & 39   & 0.22     & 10    & 5    \\
Ga   & 9.5       & 4.0    & 0.057        & 10    & 0   & W    & 0.093     & 13   & 0.071    & 15    & -123 \\
Ge    & 32.0      & 1.1    & 0.015        & 15    & 0   & Re   & 0.037     & 0.35 & 0.0019   & 15    & 20   \\
As    & 1.77      & 0.05   & 0.00067      & 30    & 0   & Os   & 0.450     & 3.9  & 0.021    & 15    & 13   \\
Se    & 20.6      & 0.090  & 0.0011       & 20    & 19  & Ir   & 0.424     & 3.5  & 0.018    & 15    & 9    \\
Br    & 3.489     & 0.076  & 0.00095      & 35    & 34  & Pt   & 0.864     & 7.1  & 0.036    & 15    & 0    \\
Rb    & 2.277     & 0.625  & 0.0073       & 10    & 4   & Au   & 0.149     & 1.7  & 0.009    & 15    & 41   \\
Sr    & 7.694     & 20.4   & 0.233        & 10    & 2   & Hg   & 0.322     & 10   & 0.05     & F4    & 0    \\
Y     & 1.504     & 3.99   & 0.045        & 10    & -8  & Tl   & 0.141     & 3.5  & 0.017    & 40    & 0    \\
Zr    & 3.669     & 9.72   & 0.107        & 10    & -8  & Pb   & 2.67      & 155  & 0.75     & 20    & 3    \\
Nb    & 0.283     & 0.605  & 0.0065       & 15    & -9  & Bi   & 0.111     & 2.5  & 0.012    & 30    & 0    \\
Mo    & 0.934     & 0.0500 & 0.00052      & 15    & 0   & Th   & 0.0293    & 77.6 & 0.33     & 10    & -2   \\
Ru    & 0.631     & 0.0070 & 0.000069     & 15    & 29  & U    & 0.0078    & 20.6 & 0.086    & 10    & 1    \\
\hline
\end{tabular}
\par\medskip \raggedright
Concentrations reported in ppmw or ppbw by weight (except for elements reported in wt\%) and in $\mu$ moles = 10$^{-6}$ moles and nmoles = 10$^{-9}$ moles. The column $\pm$ is a subjective judgment of uncertainty (see the text for more details). $\Delta$ values are \% differences relative to the estimate reported in \cite{mcdonough1995}.
F = factor, F3 means estimate known to be within a factor of 3.  See the text for details on the composition of the CI chondrite (reported in wt\% or ppmw by weight).
\end{table}

\subsection{Assessment of uncertainty}

Assessment of uncertainties for the composition of the bulk silicate Earth is a challenge. One cannot conduct a full and rigorous error propagation of all of the uncertainties as is carried out in physics experiments. This is because in the inaccessible Earth there are many potential unknowns that we cannot fully assess (e.g., deep mantle domains: D``, LLSVP, EER (early enriched reservoir \citep{boyet2005142nd}), etc.).  The uncertainties reported in Table \ref{tab:BSE} attempt to represent a reasonable approach to statistical and systematic uncertainties. All refractory lithophilic elements are assigned a combined statistical and systematic uncertainty of $\pm$ 10\%, with systematic uncertainties coming from determining the BSE enrichment factor relative to CI chondrite and uncertainties in the reference values of CI carbonaceous chondrite. The uncertainties in the main elements (O, Fe, Mg, Si) are set at $\pm$ 10\% and are driven by systematics and uncertainties in unknowns in the deep mantle, with the Mg number = 0.89 $\pm$0.o1, D$_{Si}$ value = 1, D$_{Fe}$ value $\sim$0.95. Another example of systematic uncertainty is with the estimate for P in the BSE. Initial estimates were based on a constant P/Nd value in a wide range of mantle-derived lavas (basalts, komatiites, kimberlites) \citep{mcdonough1985isotopic}. However, differences in the absolute value of P/Nd (estimates range from 55 to 75) are due to a calibration issue with the assumed concentration of P in the standard reference material. This concern remains, even though it is recognized that P/Pr (280 to 350) is a preferred ratio (that is, a plot of log P vs log Pr with a slope 1.0 line). Other uncertainties (e.g., major and minor elements) are associated with the data reported in the literature and the assumption of whole mantle convection.

\subsection{The compositional model of CI Chondrite}

As in \cite{mcdonough1995}, CI chondrite is used as a reference state composition. Mantle samples (peridotites and basalts) were used to calculate the composition of the BSE and determine the enrichment factor for refractory lithophile elements (that is, $EF_{RLE}$ = 2.65 times CI chondrite). Once this $EF_{RLE}$ is established, this multiplier is used to set the concentrations for all the RLE by multiplying it by the composition of CI chondrite. This step is taken because the modeling technique used does not allow one to determine the initial abundances of the highly incompatible RLE in the mantle (see Figure \ref{fig:peridotite}). Therefore, it was again necessary to re-establish the composition of CI chondrite, emphasizing that an internally consistent data set is used for comparison with the composition of Silicate Earth.

In establishing this composition of CI, we used data for CI chondrite and other chondrites (see also Table 3 in \cite{mcdonough1995}), particularly to examine relatively constant refractory element ratios in chondritic meteorites \citep[e.g.,][]{braukmuller2018chemical}. In recent years, compositional models for CI chondrites  \citep{barrat2012geochemistry,palme2014cosmochemical,alexander2019quantitative_CC,lodders2021relative,palme2022composition} have been improved and compared with the composition of the Sun's photosphere (see Figure \ref{fig:photo}). In addition, \cite{horan2003highly} and \cite{walker2016siderophile} provided data for highly siderophile elements and W in chondrites, including CI. Table \ref{tab:BSE} reports my estimate of the composition of CI chondrites.

\subsection{Observations on the BSE model}

\cite{mcdonough1995} noted that there are two general classes of compositional models to derive BSE: pyrolitic and chondritic. The former uses data for basalts and peridotites from the mantle to build the pyrolite model \citep{ringwood1966chemical}. However, this model is underpinned by recognition that chondrites, the building blocks of the planets, provide a guide to the mixing proportions of melt (basalt) and residue (peridotite) needed to establish the initial composition of the BSE.  The latter starts with a composition model using a chondritic reference and fits the observable Earth to this model by adjusting the components to the upper mantle, lower mantle, or core \citep{javoy2010chemical}. The pyrolite model asserts that the compositional model of BSE is homogeneous (that is, no chemical layering) and is supported by seismological evidence of mass exchange between the upper and lower mantle \citep{fukao2013subducted,french2015broad}. The chondrite model concludes compositional layering in the mantle because it is forced to have chemical distinctions between the upper and lower mantle, with the boundary being the well-defined seismic discontinuity at 660 km depth. Similar to \cite{mcdonough1995}, I conclude that BSE has a pyrolitic composition and that at the gross scale there is no compositional layering in the mantle.
At finer scales, observable heterogeneities like LLSVP and  D`` can be developed in pyrolitic models because they are small volume domains (order of $<$ 10\%) that develop by differentiation.

\begin{figure*}[h]
\centering
\includegraphics[width=0.9\linewidth]{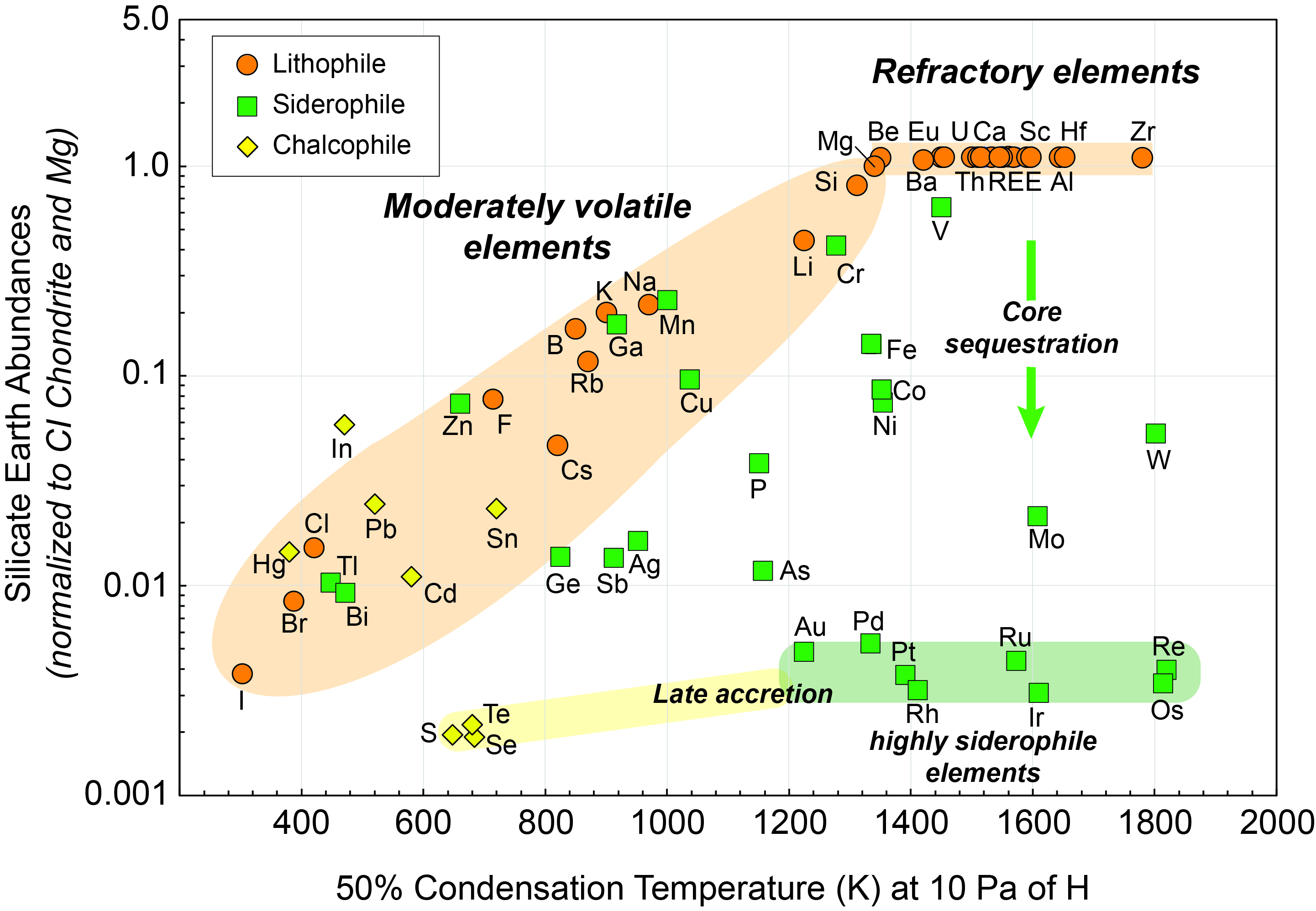}
\caption{The abundances of elements in the bulk silicate Earth normalized to Mg and Cl chondrites versus their 50\% condensation temperature ($T_{50}$). Data from Table \ref{tab:BSE} and \cite{lodders2020solar,lodders2023solar}. The orange field highlights the distribution of lithophile elements and the green and yellow fields highlight those elements that were depleted by core subtraction and later overprinted by additions to the mantle after core separation (i.e., late accretion, aka late veneer \citep{morgan1986ultramafic,walker2009highly}). Only a subset of refractory lithophile elements are identified for clarity.}
\label{fig:normal}
\end{figure*}

Figure \ref{fig:normal} shows the abundances of elements in BSE normalized with Mg and CI chondrite. The double normalization scheme compensates for the significant amount of water and CO$_2$ found in CI chondrite, relative to the Earth. Also, Mg is chosen as the normalizing element because it is the least likely of the four most abundant elements to be in the core. There is a $\sim$10\% enrichment in refractory lithophile elements and a uniform decrease in the abundances of moderately volatile lithophile elements with decreasing condensation temperature. This trend may reflect the nature of the nebular material in the planetary feeding zone of the proto-Earth at 1 AU. The chondritic pattern of highly siderophile elements and chalcogens is interpreted as a late addition to silicate Earth, post-core formation (aka late veneer) \citep{morgan1986ultramafic,walker2009highly}. 

By determining the absolute concentration of one or more refractory lithophile elements in the bulk silicate Earth (e.g., Ca, Al, Yb) the absolute abundances of the 27 refractory lithophile elements (Be, Al, Ca, Sc, Ti, Sr, Y, Zr, Ba, REE, Hf, Ta, Th, and U) can be determined. This is because chondrites have relatively constant chondritic ratios (i.e., ratios that vary by $<$15\%) for refractory lithophilic elements. Moreover, the bulk silicate Earth has been shown to have a chondritic proportion of these elements \citep{mcdonough1995} and isotopes \citep{willig2019earth,stracke2025geochemical}.

\begin{table}[h]
\caption{Major and minor oxides in the bulk silicate Earth (BSE)}
\label{tab:oxide}
\centering
\begin{tabular}{c|cc||c||ccc}
\\[-2ex]
\hline
      & Turcotte & Palme  &  this & Jagoutz & Lyubetskaya  & Javoy \\
      & \& Schubert & \& O'Neill  & study & et al & \& Korenaga  & et al \\
\hline
SiO$_2$  & 43.9        & 45.39 & 44.92      & 45.14   & 44.95 & 49.82       \\
TiO$_2$  & 0.32        & 0.211 & 0.200      & 0.217   & 0.158 & 0.12        \\
{\color[HTML]{0432FF} \textbf{Al$_2$O$_3$}$^*$} & {\color[HTML]{0432FF} \textbf{6.71}} & {\color[HTML]{0432FF} \textbf{4.50}} & {\color[HTML]{0432FF} \textbf{4.19}} & {\color[HTML]{0432FF} \textbf{3.97}} & {\color[HTML]{0432FF} \textbf{3.52}} & {\color[HTML]{0432FF} \textbf{2.42}} \\
Cr$_2$O$_3$ & --       & 0.321 & 0.334      & 0.400   & 0.385 & 0.34        \\
MnO   & --   & 0.136 & 0.135      & 0.130   & 0.131 & --          \\
FeO   & 8.10 & 8.10  & 8.05       & 7.82    & 7.97  & 8.84        \\
NiO   & --   & 0.208 & 0.219      & 0.236   & 0.252 & 0.22        \\
MgO   & 35.7 & 36.76 & 37.81      & 38.30   & 39.50 & 36.50       \\
CaO   & 5.36 & 3.65  & 3.27       & 3.50    & 2.79  & 1.87        \\
Na$_2$O  & --          & 0.349 & 0.360      & 0.326   & 0.298 & --          \\
K$_2$O   & 0.037       & 0.031 & 0.031      & 0.031   & 0.023 & --          \\
P$_2$O5  & --          & 0.020 & 0.021      & --      & 0.015 & --          \\
\hline
total & 100.0          & 99.7  & 99.6       & 100.1   & 100.0 & 100.1      \\
    &    &    &  HPE  &    &    &    \\
K (ppmw)   & 310      & 260      & 260      & 260      & 190      & 144      \\
Th (ppbw)  & 121      & 84.9     & 77.6       & 94       & 62.6     & 42       \\
U (ppbw)   & 31       & 22.9     & 20.6     & 26       & 17.3     & 12       \\
TW        & 29.3     & 21.6     & 20.0     & 23.8     & 16.1     & 11.2     \\
\hline
\end{tabular}
\par\medskip \raggedright
$^*$ ordering of BSE models according to their level of refractory lithophile element (RLE) enrichment. HPE: Heat Producing Elements. Oxides reported in wt\%. \\ 
Data sources: \cite{turcotte2002geodynamics}, \cite{palme2014cosmochemical}, \cite{jagoutz1979abundances}, \cite{lyubetskaya2007chemical}, \cite{javoy2010chemical}. \cite{turcotte2002geodynamics} compositional model fit from proportions of RLE and assumed values for SiO$_2$, MgO and FeO contents.
\end{table}

A range of compositional models for the BSE is presented in Table \ref{tab:oxide}. I have calculated the bulk compositional model of \cite{turcotte2002geodynamics} using its high levels of refractory elements (that is, Th and U are reported as 120 ppbw and 30 ppbw, respectively). The abundance of Th and U sets the enrichment level of the refractory lithophile elements, and from that the Ca, Al, and Ti content of BSE can be calculated. Taking an average Fe content of 6.3 wt\% for the BSE, this sets the MgO content, given an Mg number of 0.89 $\pm$ 0.01, and then use SiO$_2$ to bring the total sum to 100\%. The content of refractory oxide Al$_2$O$_3$, highlighted in blue, varies by almost a factor of three between models, as does heat production (TW). Both the \cite{turcotte2002geodynamics} and \cite{javoy2010chemical} models specify that the mantle is layered with an upper mantle that is pyrolite-like in composition \citep{mcdonough1995}.

The compositional model most depleted in refractory lithophile elements is that reported by \cite{javoy2010chemical}, the Enstatite Earth model. The strength of this model is that Earth and enstatite chondrites have the same isotopic composition of O, a significant observation given that half of the planet's atoms are O. However, the reduced and sulfur-rich compositions of enstatite chondrites are distinctly different from that of Earth. Another model with a depleted composition for highly incompatible elements, including refractory lithophile elements, is that of \cite{oneill2008collisional} (not shown in the Table). This model invokes an early Earth collisional erosion process to explain both the putative anomalous $^{142}$Nd isotopic compositions and the low $^4$He flux at mid-ocean ridges. In this model \cite{oneill2008collisional} have about 10 ppbw U and Th/U of $\sim$4. These two papers achieve their compositional models by different processes, but characteristically end up having limited radioactive power left in the mantle, given a continental crust with 7 TW of power.

The range of proposed compositional models reported in Table \ref{tab:oxide} can be grouped into three types, with BSE having 10, 20 or 30 ppbw U, which are roughly equivalent to 10, 20 and 30 TW of radiogenic power in Earth \citep{Sramek2013}. Considering that the continental crust has approximately 7$\pm$1 TW of power, this translates to the mantle having 3, 13, and/or 23 TW of residual power. Hence, given the range of uncertainties in these models, there is an approximate order of magnitude uncertainty in the community-wide assessment of the radiogenic power driving the Earth's mantle dynamics.

Assuming a mid-range BSE model with 20 ppbw U, the continental crust contains about 30 to 40\% of the planet's budget of K, Th and U and up to 50\% of the most highly incompatible elements \citep{rudnick2014composition}. Similar results would be obtained using other continental crustal models \citep{Kelemen2025crust,sammon2021geochemical}. To explain these crustal abundances requires processing much more than the top 27\% of the mantle mass (i.e., the mantle above 660 km depth). Therefore, compositional models for the continental crust require that at least some fraction of the lower mantle contributed to the formation of the continents. This observation is consistent with observations derived from global tomography (as described earlier), and thus also supports the concept of whole mantle convection.

\subsection{Th/U and \texorpdfstring{$^{238}$U/$^{204}$Pb}{238U/204Pb}}
The Earth’s engine is powered by an unknown proportion of primordial and radiogenic power with Th (40\%), U (40\%) and K (20\%) producing virtually all of its radiogenic heat; the rest of the decay chains (Rb-Sr, Sm-Nd, Lu-Hf, etc.) contribute the remaining 0.5\% of radiogenic heat \citep{ruedas2017radioactive,mcdonough2020radiogenic}. Currently, we are measuring the flux of geoneutrinos (see Section \ref{nu}) from the Earth to establish how much and what proportion of radiogenic fuel remains in the Earth \citep{mcdonough2023neutrino}. The analyses here represent an independent cross-check and constraint on the absolute amount of Th and U and the BSE's ratio of Th/U.

Using the compositional model presented in Table \ref{tab:BSE}, the Earth started with 56 ppbw U and 99 ppbw Th and a Th/U ratio of 1.8, which has since increased to 3.8. Using 150,000 samples of igneous, metamorphic, and sedimentary rocks and sediments, \cite{wipperfurth2018earth} found complementary elemental ratio and time-integrated Pb isotopic ratios of Th/U for rocks from the continental crust and mantle, and their values bracketed those seen in primitive meteorites. The mobility of U in fluids is documented by the large variation in Th/U in surface rocks, including U deposits, whereas the well constrained, time integrated $^{206^*}$Pb/$^{208^*}$Pb (i.e., $K_{Pb}$) shows that this ratio experienced limited fractionation during crustal extraction. The molar ratios for the continental crust $^{CC}K_{Pb}$ = 3.95$^{+0.19}_{−0.13}$ and the modern mantle $^{MM}K_{Pb}$ = 3.87$^{+0.15}_{−0.07}$ compared to the solar system $^{SS}K_{Pb}$ = 3.89$\pm$0.15 implicate a bulk silicate Earth of $^{BSE}K_{Pb}$ = 3.90$^{+0.13}_{−0.08}$ (or Th/U =3.77 for the mass ratio) \citep{wipperfurth2018earth}.

This result is the most precise elemental ratio known for BSE and demonstrates that our commonly held assumption of chondritic proportions (i.e., Th/U = 3.77 for the mass ratio, Table \ref{tab:BSE}) for refractory lithophile elements is fully justified in this case.  Moreover, the uncertainties for Th and U in the BSE reported here and in \cite{mcdonough1995} can now be updated to $\pm$ 10\% for both elements. This constraint on the Th/U value of BSE also provides an opportunity to cross-check our result with that from geoneutrino studies, which is an independent measure of the planetary molar ratio. Finally, these data place a significant constraint on the potential for Th and U partitioning into the Earth's core. If U is argued to be in the Earth's core, then there must also be Th in the core and in chondritic proportions in order to maintain the chondritic Th/U value in the BSE.

Finally, assuming an initial Pb isotopic composition for Earth as that of Nantan troilite \citep{blichert2010solar}, a BSE with 20.6 ppbw U, a $\mu$ ($^{238}$U/$^{204}$Pb) of 8.5, and a bulk Earth with 625 ppbw Pb \citep{fischer2025earth}, we obtain an estimated core composition with 1.6 ppmw Pb. If the timing of core extraction is about 50 million years after $t_{CAI}$, then the bulk Earth has a $\mu$ of 1.2.

\subsection{Noble gases and volatile elements in the Earth} \label{n-gases}

It remains uncertain as to whether Earth's accretion was occurring in the presence of an atmosphere that was itself surrounded by a protoplanetary disk. The presence of an atmospheric envelope during accretion can set the cooling conditions for surface magma oceans. The partitioning of noble gases and volatile elements (e.g., H, C, N, and $\pm$ S) between the surface and the interior of the Earth is significantly influenced by the presence of a surrounding atmosphere. Metal-silicate partitioning during core separation under specific temperature, pressure, and redox conditions controls the core's budget of these elements. Moreover, the relative proportions of the volatile elements establish the oxidation state for the planet. These two groups of elements, volatiles and noble gases, are often considered together and are highlighted by blue fields in Figure \ref{fig:photo}.

Table \ref{tab:BSE} provides estimates and uncertainties for the abundances of H, C, and N in the BSE. Current estimates for these elements in the BSE (Table \ref{tab:HCN}) vary considerably between models, giving a sense of the degree to which abundances are known (e.g., N) and which remain poorly constrained (e.g., C) \citep[e.g.,][]{mcdonough1995,marty2012origins,hirschmann2018comparative}. The uncertainty estimates for the abundances of H, C and N are generally on the order of a factor of three. The mass of surface water is 1.48 $\times$ 10$^{21}$ kg and an ocean mass is equivalent in the mantle is about 370 ppmw H$_2$O. Table \ref{tab:HCN} shows that estimates for the amount of water (in terms of ocean masses) in the BSE differ by up to an order of magnitude. Estimates for the abundances of the volatile elements in the core are not well-constrained, and therefore estimates for the bulk Earth are poorly known. We determined the concentrations of volatile elements in the bulk Earth (BE) using the core concentration estimates reported in \cite{fischer2025earth}.
\begin{table}[h]
\caption{Volatile element concentrations in the bulk silicate Earth (BSE) and bulk Earth (BE)}
\label{tab:HCN}
\centering
\begin{tabular}{c|ccccc}
\\[-2ex]
\hline
BSE &  $^{**}$ocean     & H$_2$O     & H$^*$     & C       & N     \\
 (ppmw) &   masses & & & &\\
\hline
Marty (2012)$^\dagger$  & 10 $\pm$ 5 & 4000 ± 2000 & 400 ± 220 & 790 ± 310 & 2.5 ± 1.3 \\
Hirschman (2018) &  1.9 $\pm$ 0.2 & 710 ± 90    & 80 ± 10   & 140 ± 40  & 2.8 ± 0.6   \\
McDonough \& Sun (1995) &  1.5 -- 5 & 550 -- 1900  & 60 -- 210  & $\sim$100 & $\sim$2     \\
this study &  2.5 & 900         & 100       & 120       & 2         \\
\hline Bulk Earth & & \hspace{1cm}  \\

 (ppmw)     & ocean masses  & H$_2$O         & H$^*$         & C         & N \\
this study  &  6.5   & 2350         & 260       & 730       & 25          \\
\hline
\end{tabular}
\par\medskip \raggedright
$^\dagger$ BSE recalculated from Marty's (2012) Bulk Earth composition, assuming no H, C, or N in the core. \\
$^{**}$ number of ocean mass in the BSE; one ocean mass is equivalent to about 370 ppmw of H$_2$O. \\
$^*$ H calculated from H$_2$O. \\
Bulk Earth uncertainty estimates for H, C and N in this study are a factor of 3.
\end{table}

Importantly, the delivery and the mechanism and timing of delivery of these elements to Earth play a significant role in establishing their source(s), composition, abundance, and potential scenarios for degassing. In this regard much depends on when in-gassing occurred during planetary growth: (1) in the presence of the protoplanetary disk, (2) before or after a Moon-forming event(s), and/or (3) during final $\leq$1\% mass addition to the Earth, post-core formation. For noble gases, there are three source options (nebular, chondritic, and solar wind irradiated meteoritic material). The primordial component of noble gases are the non-radiogenic isotopes ($^3$He, $^{22}$Ne, $^{36}$Ar, $^{86}$Kr, and $^{130}$Xe); these were acquired during accretion, with losses only occurring through Jeans escape ($^3$He) and loss of the atmosphere during cataclysmic events (e.g., via collisional erosion). 

The noble gas isotopic compositions of the atmosphere and mantle are measurably different and reveal that these components originated from distinct cosmochemical sources \citep{honda1991possible,ballentine2005neon}. 
\cite{mukhopadhyay2012early,parai2015evolution} showed that the mantle, as represented by mid-ocean ridge basalts (MORB) and ocean island basalts (OIB), contains two distinct noble gas components and that these sources were differentiated at least 4.45 billion years ago. The OIB source has a neon isotopic composition that is solar, while the MORB mantle source is chondritic in composition \citep{honda1991possible,ballentine2005neon}. However, both mantle sources have chondritic $^{38}$Ar/$^{36}$Ar, $^{84}$Kr/$^{82}$Kr, and $^{124}$Xe/$^{130}$Xe \citep{raquin2009atmospheric,holland2009meteorite,caracausi2016chondritic,mukhopadhyay2012early}.  Using the combined I–Pu–Xe system, \cite{mukhopadhyay2012early} showed that the mantle sources of Iceland (OIB) and MORB evolved with different I/Xe ratios and concluded that the initial phase of Earth accretion was volatile-poor compared to the later stages of accretion, which is consistent with the observations above of depletions in Earth abundances of moderately volatile lithophile elements (Figure \ref{fig:normal}).

More recently, using krypton isotopes \cite{peron2021deep} showed that basalts from Iceland and Gal{\'a}pagos (OIBs) sample a primitive noble gas source of chondritic, carbonaceous material. They concluded that throughout accretion volatiles were being delivered to a growing Earth with some being sequestered into the core and others being subjected to atmospheric loss. They concluded that much of the noble gases in the atmosphere were delivered post-Moon-forming event(s). Moreover, \cite{peron2022krypton,peron2025chondritic} found that martian atmospheric krypton was derived from accretion of solar nebula gas after formation of the mantle and before the dissipation of the solar nebula. They also noted that the martian mantle is heterogeneous, having multiple source reservoirs of chondritic volatiles. Thus, the interior inventory of volatiles and noble gases in Earth and Mars seems to have inherited chondritic volatiles in the presence of the solar nebula in the first couple of million years after solar system formation. These chondritic volatiles could have been accreted from enstatite chondrite-like bodies \citep{piani2020earth} or carbonaceous chondrites.

Overall, the mantle records an early origin of solar Ne for the deep OIB source, and later additions of mixed solar and chondritic gases for Ar, Kr and Xe as well as for H, C, and N; this later mixture is recorded in both MORB and OIB sources. The atmosphere has a carbonaceous-like chondritic component (Ne-Ar), as well as a cometary component for the heavy gases (Kr and Xe), although the cometary contribution did not affect the atmospheric budget of H, C, nor N \citep{marty2012origins}. Much of the gas budget of the mantle was inherited early during accretion. The mantle budget of Earth for H and N is consistent with that of an enstatite chondrite source. 

Estimates for the amount of noble gases in the BSE can be determined using proxy information. \cite{arevalo2009k} determined the BSE's K/U of 13,800 $\pm$ 1300 (1 standard deviation), equating to 280 ppmw K in the BSE; \cite{farcy2020k} revised this K/U to $>$12,100 and K to $>$260 ppmw. In addition, the atmosphere $^{40}$Ar/$^{36}$Ar weighs 5.137 $\times$ 10$^{18}$ kg \citep{trenberth1981seasonal}, has 1.65 $\times$ 10$^{18}$ moles of Ar, with an initial $^{40}$Ar/$^{36}$Ar of 0.0014 \citep{begemann1976primordial}, a current atmospheric value of 298.6 \citep{lee2006redetermination}, and an average mantle $^{40}$Ar/$^{36}$Ar of 12,500 \citep{mukhopadhyay2012early}. Assuming no loss of Ar from the planet, then the BSE has 3.74 $\times$ 10$^{18}$ moles of $^{40}$Ar, 5.71 $\times$ 10$^{15}$ moles of $^{36}$Ar, and a $^{40}$Ar/$^{36}$Ar of 655. Given that the atmosphere hosts some 44\% of the $^{40}$ Ar produced during Earth's history \citep{arevalo2009k}, the mantle has 2.1 $\times$ 10$^{18}$ moles of $^{40}$Ar and 1.7 $\times$ 10$^{14}$ moles of $^{36}$Ar. The integrated production of $^4$He/$^{40}$Ar is 1.8 for the U and K contents of the BSE (Table \ref{tab:BSE}). The He content of the BSE comes from assuming equal degassing of the mantle for He and Ar and using the average MORB He isotopic composition of R/R$_A$ of 8 (where R is the $^3$He/$^4$He value of the sample and R$_A$ is that of the atmosphere), then the present-day mantle has 3.8 $\times$ 10$^{18}$ moles of $^{4}$He and 4.0 $\times$ 10$^{13}$ moles of $^{3}$He. From this, a mantle with a $^3$He/$^{22}$Ne between 3 and 5 has (8 to 13) $\times$ 10$^{12}$ moles of $^{22}$Ne. Given that the atmosphere contains more than 90\% of the BSE budget of Kr and Xe, I did not calculate these abundances.

\begin{table}
\centering
\caption{$^4$He production in the core$^\dagger$ }
\begin{tabular}{r|cccccc}
\\[-2ex]
\hline
isotope  & $^{190}$Pt     & $^{184}$Os    & $^{186}$Os    & $^{204}$Pb    & $^{209}$Bi    & $^{183}$W     \\
atomic fraction   & 0.00013   & 0.0002   & 0.0159   & 0.014    & 1.0      & 0.1431   \\
atomic mass     & 195.078   & 190.23   & 190.23   & 207.2    & 208.98   & 183.84   \\
half-life      & 4.899$\times 10^{11}$ & 5.6$\times 10^{13}$  & 2.0$\times 10^{15}$  & 1.4$\times 10^{17}$  & 2.1$\times 10^{19}$  & 6.7$\times 10^{20}$  \\
e$^{(\lambda t)}$  & 0.99356   & 0.99994  & 1.00000  & 1.00000  & 1.00000  & 1.00000  \\
ppmw of element   & 5.36      & 2.79     & 2.79     & 1.60     & 0.090    & 0.550    \\
moles of isotope  & 6.9$\times 10^{15}$   & 5.7$\times 10^{15}$  & 4.5$\times 10^{17}$  & 2.1$\times 10^{17}$  & 8.3$\times 10^{17}$  & 8.3$\times 10^{17}$  \\
initial atoms isotope & 4.2$\times 10^{39}$   & 3.4$\times 10^{39}$  & 2.7$\times 10^{41}$  & 1.3$\times 10^{41}$  & 5.0$\times 10^{41}$  & 5.0$\times 10^{41}$  \\
$^4$He atoms produced  & 2.7$\times 10^{37}$   & 1.9$\times 10^{35}$  & 4.3$\times 10^{35}$  & 2.8$\times 10^{33}$  & 7.5$\times 10^{31}$  & 2.4$\times 10^{30}$  \\
\hline
moles of $^4$He    & 4.47$\times 10^{13}$  & 3.20$\times 10^{11}$ & 7.12$\times 10^{11}$ & 4.72$\times 10^{9}$ & 1.25$\times 10^{8}$ & 3.90$\times 10^{6}$ \\
kg of $^4$He    & 1.79$\times 10^{11}$  & 1.28$\times 10^{9}$ & 2.85$\times 10^{9}$ & 1.89$\times 10^{7}$ & 5.01$\times 10^{5}$ & 1.56$\times 10^{4}$  \\
\hline

\end{tabular}

\par\medskip
$^\dagger$ Mass of the core is 1.93 $\times$ 10$^{24}$ kg.
\label{table:4He}
\end{table}

To understand the core's chemical and isotopic composition of noble gases, we look to ocean island basalts, whose sources may be affected by gases from the core. Baffin Island lavas (the proto-Iceland plume) with the highest terrestrial $^3$He/$^4$He value (R/R$_A$ = 67, \citep{horton2023highest}) are argued to have noble gases that originate from the core, as it is a place where alpha production within the Earth is minimal. The measured helium isotope ratio of Jupiter is our best estimate of the proto-solar value of ($^3$He/$^4$He = 1.66 $\pm$ 0.05 $\times$ 10$^{-4}$ (R/R$_A$ = 120); \citep{mahaffy1998galileo}). The Earth's mantle value for $^3$He/$^4$He is current assumed to be between 1.1 and 4.2 $\times$ 10$^{-5}$ \citep{peron2018origin}, with the average R/R$_A$ for MORB is 8 which is equivalent to 1.33 $\times$ 10$^{-5}$.

There are alpha-decay producers in the core. Table \ref{table:4He} lists the six siderophile isotopes that undergo alpha decay to produce $^4$He. The last three rows report the number of atoms, moles, and kg of $^4$He produced in the core over the age of the Earth. The highest amount of $^4$ He produced during the lifetime of the core comes from $^{190}$Pt, and its production is comparable to the lower limit of what might be the core's $^3$He content.  If we use the basalt's high $^3$He/$^4$He value (67) as an upper limit along with the core's content of alpha producers, then we can calculate a lower limit on the amount of He in the core, which is 1.25$\times$10$^6$ atoms $^3$He/g of core.

In evaluating the core as a possible source of mantle helium, \cite{porcelli2001core} estimated that the core has $\sim1\times$10$^{9}$ atoms $^3$He/g  or $\sim$2$\times$10$^{36}$ atoms of $^3$He based on metal-silicate partition coefficients for He \citep{matsuda1993noble} and an initial $^3$He/$^4$He of 1.66 $\times$ 10$^{-4}$. Later, high P-T experiments by \cite{bouhifd2013helium} estimated that at least 1.2$\times$10$^{11}$ atoms $^3$He/g may reside in the core, which is, 10$^5$ times greater than the lower limit calculated above. More recent studies continue to reinforce this view that the metallic core contains even greater amounts of helium \citep{bouhifd2020potential,roth2019primordial,takezawa2025formation} and is the source of some oceanic basalts with high $^3$He/$^4$He isotopic compositions. An alternative perspective was presented by \cite{li2022primitive} where they concluded that there are three orders of magnitude difference in the metal-silicate partitioning for He and Ne and that this difference would results in anomalously low $^3$He/$^{22}$Ne gas compositions, which have not been observed.

Another consideration is the question of U in the Earth's core. If U is assumed to be in the core and provides the power to drive the geodynamo, a scenario that is not supported by \cite{fischer2025earth}, and the core is the source of high R/R$_A$ basalts, then a limit can be established on the amount of U in the core. \cite{bouhifd2020potential} presented two case examples for the dissolution of helium in the core that resulted in \textit{Case I having 1.2$\times$10$^{11}$ }and \textit{Case II having 3$\times$10$^{10}$} atoms $^3$He/gram of core. If the metal-silicate partition coefficient for U is 10$^{-3}$ during core formation, then it would result in $\sim$ 21 pg/g U in the core. Table \ref{table:UandK} shows that this amount of U can produce $\sim$10$^{39}$ atoms of $^4$He over the lifetime of the core and would again result in a negligible change in the $^3$He/$^4$He value of the core. If we assume that all U predicted to be in the BSE (20.6 ppbw) were hosted in the core, this would produce 2.7$\times$10$^{42}$ atoms of $^4$He. (Adding $^{232}$Th to this calculation produces a total of 4.0$\times$10$^{42}$ atoms of $^4$He over the age of the Earth.) Adding this amount of $^4$He atoms to the core would change its R/R$_A$ value from 120 to 64 in the example of \textit{Case I} and from 120 to 16 in the example of \textit{Case II}. These calculations show that the core's potential for $^4$He production is negligible compared to the amount of helium that may have been dissolved in the metal during core formation.

Finally, it has also been suggested that the core might contain some K as an energy source to drive the dynamo \citep{corgne2007how,watanabe2014abundance}. This condition is believed to be appropriate for core formation which led \cite{watanabe2014abundance} to suggest up to 40 ppmw K in the core. \cite{roth2019primordial} proposed that in addition to He and Ne, other noble gases could be equally partitioned into the core at the earliest stage of core formation. Assume a $^3$He/$^{36}$Ar of 0.53 for the core and \cite{bouhifd2020potential} \textit{Case I} scenario for the amount of $^3$He in the core, then one has 1.2$\times$10$^{38}$ atoms of $^{36}$Ar and a $^{40}$Ar/$^{36}$Ar of 2800 in the core. The addition of a core noble gas contribution to an OIB source could have a distinctively high $^3$He/$^4$He value, but not necessarily an anomalous $^{40}$Ar/$^{36}$Ar value.

\begin{table}[h]
\centering
\caption{Model for U produced $^4$He and K produced $^{40}$Ar in the core$^\dagger$ }
\begin{tabular}{r|cc|rc}
\\[-2ex]
\hline
isotope   & $^{235}$U    & $^{238}$U    &  & $^{40}$K       \\
atomic fraction          & 0.007205  & 0.99274   &  & 0.0001167 \\
atomic mass    & 238.0289  & 238.0289  &  & 39.098    \\
half-life & 7.035$\times 10^8$ & 4.468$\times 10^9$ &  & 1.262$\times 10^9$ \\
e$^{(\lambda t)}$        & 0.011     & 0.492     &  & 0.081     \\
ppmw of element & 2.1$\times 10^{-5}$  & 2.1$\times 10^{-5}$  &  & 40  \\
moles of isotope         & 1.22$\times 10^{12}$  & 1.68$\times 10^{14}$  &  & 4.83$\times 10^{17}$  \\
initial atoms isotope    & 6.62$\times 10^{37}$  & 2.06$\times 10^{38}$  &  & 3.57$\times 10^{42}$  \\
$^4$He atoms produced    & 4.58$\times 10^{38}$  & 8.35$\times 10^{38}$  & $^{40}$Ar atoms produced  & 3.46$\times 10^{41}$  \\
\hline
moles of $^4$He  & 7.61$\times 10^{14}$  & 1.39$\times 10^{15}$  & moles of $^{40}$Ar & 5.75$\times 10^{17}$  \\
kg of $^4$He    & 3.04$\times 10^{11}$  & 5.55$\times 10^{12}$  & kg of $^{40}$Ar & 2.30$\times 10^{16}$  \\
\hline

\end{tabular}

\par\medskip
$^\dagger$ Mass of the core is 1.93 $\times$ 10$^{24}$ kg.

\label{table:UandK}
\end{table}

\section{Bulk Earth composition}
To establish the composition of the bulk Earth, we used the results given in Table \ref{tab:BSE} for the bulk silicate Earth and those reported in \cite{fischer2025earth} for the core. Therefore, the bulk Earth compositional model (Table \ref{tab:BE}) was constructed from 0.676 of the BSE and 0.324 of the core (i.e., 2/3 and 1/3, respectively).

\begin{table}[H]
\caption{Concentration of elements in the bulk Earth (BE)}
\label{tab:BE}
\centering
\begin{tabular}{lccc|lccc}
\\[-2ex]
\hline
BSE   & 10$^{-6}$ kg/kg & $\mu$moles/g & $\pm$  & BSE & 10$^{-9}$ kg/kg   & nmoles/g  & $\pm$   \\
\hline
H  & 270    & 265    & F3 & Rh & 250  & 2.5  & 15 \\
Li & 1.1    & 0.16   & 20 & Pd & 970  & 9.1  & 15 \\
Be & 0.042  & 0.0046 & 10 & Ag & 160  & 1.4  & 35 \\
B  & 0.20   & 0.018  & 30 & Cd & 120  & 1.1  & 30 \\
C  & 750    & 62     & F3 & In & 10   & 0.10 & 40 \\
N  & 30     & 1.9    & F2 & Sn & 390  & 3.3  & 30 \\
O \% & 29.7 & 18540  & 10 & Sb & 110  & 0.9 & 50 \\
F  & 10     & 0.60   & 35 & Te & 610  & 4.8  & 20 \\
Na & 1780   & 77     & 15 & I  & 70   & 0.6 & 60 \\
Mg \% & 15.2 & 6250   & 10 & Cs & 15   & 0.11 & 40 \\
Al \% & 1.48  & 547    & 10 & Ba & 4260 & 31   & 10 \\
Si \% & 15.2 & 5400   & 10 & La & 420  & 3.0  & 10 \\
P  & 1510   & 49     & 15 & Ce & 1090 & 7.8  & 10 \\
S \%  & 1.10  & 350    & 40 & Pr & 160  & 1.2  & 10 \\
Cl & 17     & 0.49    & 35 & Nd & 830  & 5.8  & 10 \\
K  & 170    & 4.4    & 20 & Sm & 260  & 1.8  & 10 \\
Ca \% & 1.59  & 398    & 10 & Eu & 100  & 0.67 & 10 \\
Sc & 10     & 0.23   & 10 & Gd & 360  & 2.3  & 10 \\
Ti & 787    & 16     & 10 & Tb & 60   & 0.41 & 10 \\
V  & 100     & 2.0    & 10 & Dy & 450  & 2.8  & 10 \\
Cr & 4550   & 88     & 15 & Ho & 100  & 0.60 & 10 \\
Mn & 1360   & 25     & 10 & Er & 290  & 1.7  & 10 \\
Fe \% & 32.6 & 5840   & 10 & Tm & 50   & 0.27 & 10 \\
Co & 920    & 16     & 10 & Yb & 290  & 1.7  & 10 \\
Ni & 18800  & 320    & 10 & Lu & 40   & 0.25 & 10 \\
Cu & 100     & 1.6    & 15 & Hf & 190  & 1.1  & 10 \\
Zn & 80     & 1.25    & 10 & Ta & 30   & 0.14 & 10 \\
Ga & 6.0    & 0.086     & 10 & W  & 190  & 1.04  & 15 \\
Ge & 20     & 0.22    & 15 & Re & 80   & 0.41 & 15 \\
As & 1.6    & 0.022     & 30 & Os & 930  & 4.9  & 15 \\
Se & 4.7    & 0.060     & 20 & Ir & 880  & 4.6  & 15 \\
Br & 0.051   & 0.0006      & 35 & Pt & 1790 & 9.2  & 15 \\
Rb & 0.42   & 0.005      & 10 & Au & 200  & 1.0  & 15 \\
Sr & 14.0     & 0.155    & 10 & Hg & 20   & 0.12 & F4 \\
Y  & 2.66    & 0.030     & 10 & Tl & 20   & 0.10 & 40 \\
Zr & 6.48    & 0.071     & 10 & Pb & 640  & 3.1  & 20 \\
Nb & 0.403   & 0.004      & 15 & Bi & 30   & 0.15 & 30 \\
Mo & 1.87    & 0.019     & 15 & Th & 52   & 0.22 & 10 \\
Ru & 1.31    & 0.013     & 15 & U  & 14   & 0.058 & 15 \\
\hline
\end{tabular}
\par\medskip \raggedright
Concentrations reported in ppmw or ppbw by weight (except for elements reported in wt\%) and in $\mu$moles = 10$^{-6}$ moles and nmoles = 10$^{-9}$ moles. The $\pm$ column is a subjective judgment of uncertainty in \% (see text for details). F = factor, F3 means estimate known to within a factor of 3.
\end{table}

\section{Powering the Earth's Engine} \label{nu}

After being rolled off the factory floor, the assembled Earth has its complete fuel complement to drive its dynamics. Once this fuel runs out, the planet can no longer generate a magnetic field or drive plate tectonics, volcanism, and mantle convection. 

Our rocky planet has two sources of fuel: primordial, the kinetic energy from assembling the planet, and nuclear, the energy from the heat produced during natural radioactive decay. The current surface heat flux is 46 $\pm$ 3 TW (terrawatts, 10$^{12}$ watts) \citep{jaupart2015temperatures}. Models vary widely in terms of how much primordial and nuclear fuel remains inside Earth. In general, it is agreed that the continental crust contributes 7 $\pm$1 TW of radiogenic power \citep{rudnick2014composition,mcdonough2023neutrino}. Although many assume that we know the total abundances and distribution of radioactive heat producing elements in Earth (i.e., HPE, U, Th, and K), current estimates for mantle heat production vary by an order of magnitude (2 to $\geq$20 TW of radiogenic power). Thus, we lack sufficient constraints on the thermal evolution of the planet. As discussed in \cite{fischer2025earth} and in Section \ref{n-gases} that U and K are potential sources of $^4$He and $^{40}$Ar, respectively, in the core.

For the last two decades neutrino physicists have been detecting low-energy (MeV) electron antineutrinos being emitted from the Earth (geoneutrinos) that are produced via beta-minus radioactive decay of U and Th \citep{araki2005experimental}. Collectively, the flux of geoneutrinos measured at detectors reveal the radiogenic power that drives the Earth's engine \citep{mcdonough2023neutrino}. To date, we have flux measurements from Japan and Italy, a result from Canada is forthcoming in 2025, and in 2026 we expect a result from the JUNO experiment in China.

The geoneutrino flux measured in Japan using a 1 kiloTon liquid scintillation detector (KamLAND) \citep{abe2022abundances} and in Italy using a 0.3 kT detector (Borexino) \citep{agostini:2020,sammon2022quantifying} reveal that the mantle produces $\sim$13 $\pm$ 8 TW and the continental crust generates $\sim$7 $\pm$ 1 TW for a total of 20 $\pm$ 8 TW of radiogenic power. The remaining surface flux of power comes from non-radiogenic sources and is $\sim$26 TW. This latter flux likely represents subequal contributions of primordial power from the core-mantle boundary (CMB) and mantle. 

Geoneutrino measurements will define with increasing accuracy and precision the compositional model of the Earth, which in turn will place tight constraints on its thermal evolution, and independently confirm (or not) the planet's chondritic ratio of Th/U. A combined data analysis using KamLAND and Borexino geoneutrino experiments affirms that the BSE has proportions of refractory lithophile elements at 2.5 to 2.7 times those of CI carbonaceous chondrites.

Converting the measured geoneutrino flux into a global value of TW of power requires that we accurately and precisely define local geological models of the lithosphere (crust and mechanically coupled mantle) that surround these detectors. As noted in \cite{araki2005experimental}, approximately 50\% of the flux signal is contributed by this region. The flux contribution is also predicted from the mantle ($\sim$25\%) and the distant lithosphere ($\sim$25\%), but these are less critical as the detector sensitivity drops of at \textit{1/r$^2$}, where $r$ is the separation distance between the detector and the decaying atom.

Here, we highlight the importance of accurately defining a local geological model. The initial geological survey surrounding the Italian Borexino detector by \cite{coltorti2011u} did not incorporate a signal from the voluminous alkaline to ultrapotassic rocks of the Tuscan province that covers about 1/3 of the western part of the near field region (i.e., closet 250 km). Working with data from \cite{coltorti2011u}, \cite{agostini:2020} reported the final result of the Borexino experiment. They reported a radiogenic power for the mantle of 30 TW and for the overlying lithosphere of 8 TW, but with large uncertainties. Their estimate predicted an Earth with 38 ppbw U and 143 ppbw Th. In contrast, \cite{mcdonough1995} estimate that Earth has 20 and 75 ppbw U and Th, respectively. \cite{sammon2022quantifying} included the Tuscan magmatic rocks with their high K, Th and U contents in their analyzes, remodeled the geoneutrino flux data, and concluded that the Borexino result strongly favors an Earth with $\sim$20TW present-day total radiogenic power.

The Canadian Sudbury Neutrino Observatory (SNO+, 1 kT), situated in the Archean craton, has been measuring the local geoneutrino flux since April 2022 and is likely to report their measurement later this year. The 20 kiloton Chinese JUNO neutrino detector is currently being filled and is scheduled to begin counting in August 2025. Larger than all previous detectors, JUNO will receive more geoneutrino counts in one year than all previous experiments collectively have detected over their lifetime.  A mobile ocean bottom detector is being developed to map out chemical heterogeneities in the mantle. Therefore, the future is bright for continued studies in neutrino geoscience. 

At the 1$\sigma$ level, the geoneutrino data reject bulk Earth compositional models that propose high concentrations of refractory lithophile elements ($>2.5 \times$ CI chondrite), including those with high Th and U contents, as suggested by \citep{turcotte2002geodynamics}, and favor models that have bulk Earth compositions of $1.9 \pm 0.2 \times$CI chondrite \citep{yoshizaki2021earth}.  In addition, these data reveal that there is 7 to 8 mode\% davemaoite (i.e., Ca-perovskite) and $\sim$17 mode\% ferropericlase in the lower mantle. These numbers are consistent with a pyrolite compositional model for the BSE and a relatively uniform major element composition throughout the mantle, with smaller domains of chemical heterogeneities (e.g., recycled oceanic lithosphere and seismically recognized domains of low velocity \citep{cottaar2016morphology}).

In the future there is the potential to deploy an ocean-going neutrino detector that can map the chemical variation in U and Th in the mantle \citep{Sramek2013,mcdonough2023neutrino}. Efforts to fund these studies continue. Moreover, collaborations between geologists and particle physicists are also leading to greater insights in other areas, including studies of neutrino oscillations that can reveal differences in the electron density of the Earth's interior, which is a function of its hydrogen content, and neutrino absorption that can reveal details associated with absolute density and density contrasts (e.g., core-mantle boundary).

\section{Earth's accretion age and thermal history} \label{thermal}
\begin{figure*}[h]
\centering
\includegraphics[width=0.8\linewidth]{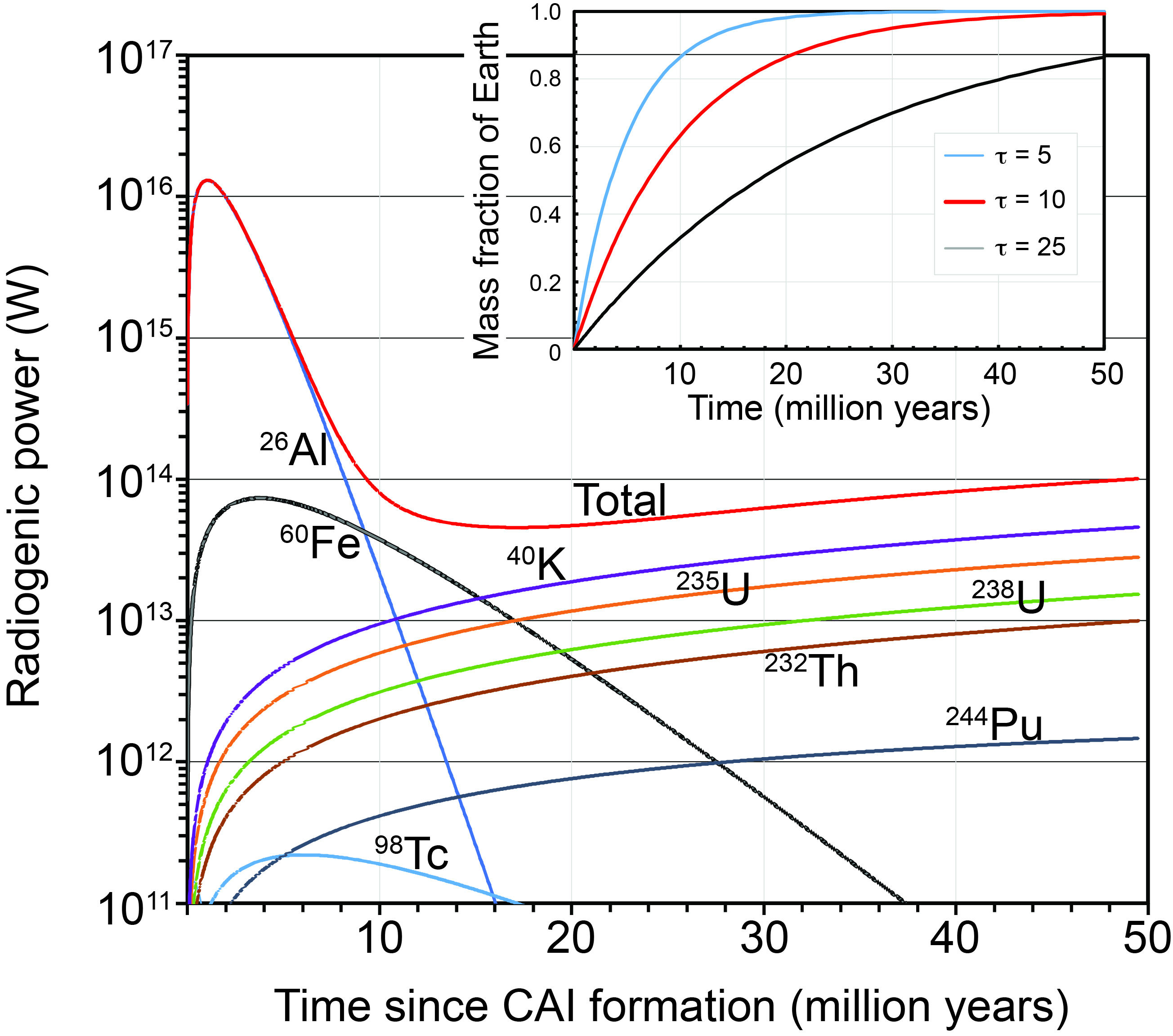}
\caption{The radiogenic heat contributions over the first 50 million years of Earth's history, including that from short‐lived radionuclides \citep{mcdonough2020radiogenic}. Inset diagram shows a series of exponential
growth curve for three different tau accretion ages. A $\tau_{accretion}$ age of 10 million years is assumed for the Earth's thermal model \citep{yin2002short}; however, the absolute $\tau$ value is not known.}
\label{fig:T-history}
\end{figure*}

Describing the age of the Earth is not easy, as there is no unique moment that defines its "birth"; for example, even today $>10^7$ kg of astromaterials accrete onto the Earth every year \citep{love1993direct}. Therefore, physically, we use the time constant $\tau_{accretion}$ (tau accretion) as a reference age for the Earth, which represents when the Earth grows to about 62\% ($\tau_{accretion}$=1-(1/e)) of its final mass. Accretion ages for the parent bodies of NC iron meteorites range from as old as CAIs ($t_{CAI}$, which is taken as the start of the solar system, 4.567 Ga; \citep{amelin2002lead,bouvier2010age,connelly2012absolute}) to up to 2 million years after $t_{CAI}$ \citep{hilton2022chemical}. \cite{dauphas2011hf,tang201460fe} determined a $T_{50}$ accretion age for Mars of 1.9$^{+1.7}_{-0.8}$ million years, a body that is 10\% of Earth's mass. Planetesimals and planets grew early and rapidly in the presence of protoplanetary disk. However, it is not known how fast growth continued for the larger rocky planets, Earth and Venus. Taking a simple growth model (M$_{(t)}$/M$_{final}$ = 1-exp$^{(-t/\tau)}$) for the Earth and a $\tau_{accretion}$ age of 10 million years \citep{yin2002short}, the growth of the planet effectively continues for up to 50 million years after $t_{CAI}$ (Figure \ref{fig:T-history} inset). 

Currently there is considerable discussion about the specific models of planetary growth in the solar system (i.e. pebble accretion versus oligarchic accretion; repeated giant impacts between planetary embryos after the protoplanetary disc stage, etc.) \citep{greenberg1978planetesimals,kokubo1998oligarchic,olson2023hafnium,morbidelli2025did,johansen2024comment}. This topic will not be covered here, as it continues to provoke strong responses in discussions despite the many under-constrained variables. However, scientific history reveals parallels between this debate and the 1980 debate surrounding whole-versus-layered mantle convection. Both topics are vigorously discussed, both involve testing a considerable number of adjustable parameters in computational modeling, and both perspectives bring into focus important constraints from chemical and isotopic studies. It is worth noting that the discussion on whole-versus-layered mantle convection continues today some 40 years later, with some interesting and much more mature ideas on the form of mantle convection.

Given constraints from geoneutrino studies and a model for Earth's growth, we can construct a model for planet heating from radioactive decay of short- and long-lived species (Figure \ref{fig:T-history}). These calculations show that despite their short half-lives (0.7 and 2.6 million years, respectively), $^{26}$Al and $^{60}$Fe contribute a significant amount of heating to the planet over its first 10 million years (Figure \ref{fig:T-history}). During this time frame, the growing Earth probably experienced temperatures that exceeded its solidus (and probably its liquidus, too) for periods of time.  Estimates of a temperature increase greater than 3000 K ($\Delta$T) in $<$1 million years ($\Delta$ t) are likely if we compare the specific power $h$ (W/kg) to C$_P$$\Delta$T/$\Delta$t, assuming a planetary value of 1,000 J kg$^{−1}$ s$^{−1}$ for the specific heat capacity (C$_p$). This temperature increase is sufficient to induce melting and enhance the effectiveness of metal‐silicate fractionation.

One can also ask how long will the Earth power supply last? After 5 billion more years of evolution, some $\sim$10 billion years after the start of the solar system, the Earth will still have about 10 TW of radiogenic power (essentially from $^{238}$U and $^{232}$Th); however, it will probably have used up most of its primordial power.

\section{Future prospects}
There are prospects for future discoveries, particularly in terms of placing significant constraints on the amount of H in the Earth. 

Neutrino experiments are expanding their applications in geosciences with neutrino oscillation \citep{rott2015spectrometry,winter2016atmospheric} and neutrino absorption \citep{kumar2021validating,upadhyay2023locating} studies. High-energy particles produced in the galaxy and beyond can interact with the Earth's atmosphere and produce high-energy cosmic-ray-air collisions, including downgoing neutrinos that will pass through the Earth and be detected as they leave the planet. Neutrino oscillation studies use some of these energetic particles to sense the electron density state of the interior of the Earth (Z/A variation with depth). The KM3NeT, Orca, P-One, and Trident experiments \citep{margiotta2014km3net,ye2023multi,malecki2024behalf} will collect 2 to 8 GeV neutrinos to sense the electron density state of the Earth as they traverse different cross sections of the planet's interior depending on the incident angle of transit and detection. These Z/A estimates can be tested with high-precision oscillation data to evaluate the chemical composition of the Earth, specifically testing models for the hydrogen content of the core and/or the Mantle Transition Zone, since hydrogen uniquely has a Z/A value of 1, while all other elements have a value of $\sim$0.5. Neutrino absorption studies use $\geq$10 TeV neutrinos to sense the state of mass density of the interior of the Earth (e.g., core-mantle boundary) and will be capable of revealing absolute density and depths of strong density contrasts in the Earth's interior.

\section*{Acknowledgments}
I thank the early reviews by Rich Walker, Rick Arevalo, FangZhen Tang, Bill White and Ming Tang for their thorough reading and constructive comments. These reviews helped me greatly. I gratefully acknowledge NSF support (EAR2050374).

\bibliographystyle{elsarticle-harv}
\bibliography{ref}
\end{document}